\documentclass[prx,twocolumn,amsmath,amssymb,superscriptaddress,floatfix,nofootinbib,aps]{revtex4-2}
\usepackage{babel}
\usepackage{amsmath}
\usepackage{amsfonts}
\usepackage{graphicx}
\usepackage{dcolumn}
\usepackage{bm}
\usepackage{amsthm}
\usepackage{algorithm}
\usepackage{algpseudocode}
\usepackage[cmyk]{xcolor}
\usepackage[colorlinks,bookmarks=false,citecolor=magenta,linkcolor=magenta,urlcolor=magenta]{hyperref}
\usepackage{braket}
\usepackage{orcidlink}

\begin{document}
\preprint{APS/123-QED}
\title{Entanglement, anti-flatness, and nonlocal nonstabilizerness:
a unified perspective from entanglement spectrum}

\author{Lei-Yi-Nan Liu \orcidlink{0009-0005-3072-3650}}
 \affiliation{%
 School of Physics, Beihang University, Beijing 100191, China
}

\author{Jian Cui \orcidlink{0000-0001-6643-7625}}
 \email{jiancui@buaa.edu.cn}
 \affiliation{%
 School of Physics, Beihang University, Beijing 100191, China
}

\date{\today}

\begin{abstract}
Entanglement and nonstabilizerness capture distinct aspects of quantum
complexity, yet their relation through the entanglement spectrum remains
poorly understood. Here we develop a spectral framework for bipartite
nonlocal nonstabilizerness. We introduce a generalized anti-flatness and
derive upper and lower bounds on the nonlocal stabilizer R\'enyi entropy
(SRE) in terms of R\'enyi entanglement entropy and spectral
non-uniformity. We further reduce the local-unitary optimization to a
single-unitary problem and prove that the ordered Schmidt reference state
is a local minimum of the SRE for integer $\alpha\geq2$. Applying these
results to exponentially and algebraically decaying spectra reveals
parametrically distinct relations between entanglement and nonlocal
nonstabilizerness. For broad spectra, we introduce a dyadic-shell
sandwich construction that determines the asymptotic SRE scaling. At
one-dimensional conformal critical points, it yields a universal
hierarchy of double-logarithmic scaling laws for integer $\alpha\geq2$. 
Our spectral bounds and dyadic-shell sandwich construction provide general tools for analyzing nonlocal SRE, offering a flexible framework that can be applied to a wide range of entanglement spectra in quantum many-body systems.
\end{abstract}
\maketitle

\textit{\textbf{Introduction}.---} 
Entanglement and nonstabilizerness characterize two distinct forms of
quantum complexity. 
Entanglement quantifies nonlocal quantum correlations and plays a
central role in the classification of quantum phases, critical
phenomena, quantum thermalization, 
and tensor-network descriptions of many-body states \cite{Horodecki2009,Osborne2002,Tantivasadakarn2024,Chen2010,XiaoGang2019,Broholm2020,Abanin2019,Kaufman2016,Brenes2020,Vidal2003,Sauerwein2019,Cirac2021}.
Nonstabilizerness, or magic, instead quantifies the departure from the
stabilizer manifold and provides the resource that promotes Clifford
circuits to universal quantum computation \cite{Gottesman1998,Bravyi2005,Howard2017,Veitch2012}.
The two resources are fundamentally different. Highly entangled
stabilizer states can carry no magic, while product states may possess
substantial magic originating entirely from local basis choices.
Understanding how entanglement and magic are related is therefore an
important step toward characterizing the complexity of quantum
many-body systems~\cite{Liu2022}. 

The stabilizer R\'enyi entropy (SRE) provides a convenient measure of
magic for many-body systems, as it is Clifford invariant and can be
expressed directly in terms of moments of Pauli expectation
values~\cite{Lorenzo2022}. It has been widely studied in quantum many-body
systems~\cite{Haug2023MPS,Tarabunga2023,Tarabunga2024critical,Tarabunga2024,
Falcao2025,Russomanno2025,Zhenyu2026,Grabarits2026,Maity2026,Odavic2025,
Turkeshi2025,Szombathy2025,Haug2025,Niroula2024,Smith2025,Hallam2026,
Tirrito2025,Tirrito2025transport,Sierant202603}, and is closely related to
the Pauli spectrum~\cite{Turkeshi2025prb}. To isolate the
nonstabilizerness irreducibly encoded in bipartite correlations, one can
minimize the SRE over local unitaries, although this leads to a highly
nonconvex optimization problem~\cite{Qian2025,Cao2025}. For fermionic
Gaussian states, restricting the optimization to local Gaussian unitaries
yields an exact solution at $\alpha=2$ in terms of the reduced Majorana
covariance spectrum~\cite{Iannotti2026,Collura2026}. More generally, a
Schmidt-spectrum construction has been proposed as an analytical candidate
for the nonlocal SRE, supported by exact low-rank results and extensive
numerical optimizations~\cite{Xiao2026,Fabio2026,Gianpaolo2026,Liu2026}.
Here we provide further evidence by proving that, for integer
$\alpha\geq2$, the ordered Schmidt reference state is a local minimum of
the SRE, although its global optimality remains open. 
These developments suggest that the entanglement spectrum (ES) contains
the essential information governing nonlocal nonstabilizerness and raise
the question of which spectral features are responsible for it. The amount
of entanglement alone is insufficient, so the detailed geometry of the ES
must play an essential role. One natural characterization is the
\emph{anti-flatness}, which quantifies deviations of the Schmidt spectrum
from uniformity and has been closely connected to
nonstabilizerness~\cite{Tirrito2024,Iannotti2025,Barbara2026}, as well as
widely studied in quantum many-body systems~\cite{Ebner2026,Zhang2026,
Robin2026,Odavic2025,Turkeshi2023}. However, spectral non-uniformity is not
the whole story. A broad spectrum can spread its Schmidt weight over many
rank scales even when its anti-flatness becomes small, suggesting that
nonlocal nonstabilizerness is sensitive to both deviations from flatness
and the distribution of Schmidt weight across logarithmic rank scales.

In this work, we develop a unified spectral framework for entanglement,
generalized anti-flatness, and nonlocal nonstabilizerness. We establish
universal upper and lower bounds on the SRE measure of nonlocal
nonstabilizerness directly from the ES, with the R\'enyi entanglement 
entropy (EE) providing the upper bound and the generalized anti-flatness
providing the lower bound. We further show that, for integer
$\alpha\geq2$, the ordered Schmidt reference state is a local minimum of
the SRE under local-unitary transformations. We apply these results to
exponentially and algebraically decaying spectra, revealing distinct
relations between entanglement, spectral non-uniformity, and nonlocal
nonstabilizerness. We further introduce a dyadic-shell description that
captures the distribution of Schmidt weight across logarithmic rank 
scales and provides a unified treatment of broad entanglement spectra.
Applied to the Calabrese--Lefevre spectrum, this framework yields a
universal hierarchy of double-logarithmic scaling laws for the nonlocal
SRE at one-dimensional conformal critical points. Our results identify
generalized anti-flatness and dyadic-shell weight distributions as
complementary spectral structures governing nonlocal nonstabilizerness.

\textit{\textbf{Bounds on nonlocal SRE}.---}
We begin by introducing the stabilizer R\'enyi entropy (SRE), a widely
used measure of quantum nonstabilizerness~\cite{Lorenzo2022}.
For an $n$-qubit pure state $\ket{\psi}$, the $\alpha$-SRE is defined as
\begin{equation}
    M_{\alpha}(\ket{\psi})
    =
    \frac{1}{1-\alpha}
    \log_2
    \sum_{P\in\mathcal{P}_n}
    \Xi_P^{\alpha}(\ket{\psi})
    -\log_2 d ,
    \label{definition}
\end{equation}
where $\mathcal{P}_n$ denotes the set of all $n$-qubit Pauli strings
with positive phase,
$d=2^n$, and $\Xi_P(\ket{\psi})=d^{-1}\bra{\psi}P\ket{\psi}^{2}$. 
The SRE is Clifford invariant and can be expressed directly in terms
of moments of Pauli expectation values, making it particularly useful
for analytical and numerical studies of many-body nonstabilizerness. 
We now consider a bipartition of $\ket{\psi}$ into subsystems $A$ and
$B$. Its Schmidt decomposition can be written as
$\ket{\psi}_{AB}=\sum_{i=0}^{\chi-1}\sqrt{\lambda_i}\ket{a_i}_A\ket{b_i}_B$, 
where $\chi$ is the Schmidt rank and
$\sum_i\lambda_i=1$. 
We arrange the nonzero Schmidt coefficients in descending order,
$\lambda_0\geq\lambda_1\geq\cdots\geq\lambda_{\chi-1}>0$.
The nonlocal $\alpha$-SRE is obtained by minimizing the SRE over local
unitary transformations,
\begin{equation}
    M_{\alpha}^{\mathrm{NL}}(\ket{\psi}_{AB})
    =
    \min_{U_A\otimes U_B}
    M_{\alpha}
    \left[
    (U_A\otimes U_B)\ket{\psi}_{AB}
    \right],
\end{equation}
which removes the nonstabilizerness associated purely with local basis
choices and retains the contribution irreducibly encoded in the
bipartite correlations \cite{Qian2025,Cao2025}.

To express the nonlocal SRE directly in terms of the entanglement spectrum, 
we consider the Schmidt reference state 
$\ket{\tilde{\psi}}=\sum_{x=0}^{d-1}\sqrt{\lambda_x}\ket{x}_{\tilde A}\ket{x}_{\tilde B}$, 
where the Schmidt coefficients are arranged in descending order.
For a spectrum of Schmidt rank $\chi$, we choose
$m=\lceil\log_2\chi\rceil$ and $d=2^m$, and pad the spectrum with zeros
so that $\lambda_x=0$ for $\chi\le x\le d-1$. The auxiliary subsystems
$\tilde A$ and $\tilde B$ therefore provide a canonical computational-basis
encoding of the original nonzero Schmidt spectrum~\cite{Fabio2026,Gianpaolo2026,Xiao2026,Liu2026}.
Motivated by recent works suggesting that the nonlocal SRE is determined
by the Schmidt spectrum and can be obtained from such a canonical
representative~\cite{Xiao2026,Gianpaolo2026,Fabio2026,Liu2026},
we identify
\begin{equation}
    M_{\alpha}^{\mathrm{NL}}(\boldsymbol{\lambda})
    :=M_{\alpha}(\ket{\tilde{\psi}})
    =\frac{1}{1-\alpha}\log_2 \zeta_\alpha ,
    \label{NL_definition}
\end{equation}
where
$\zeta_\alpha=(1/d)\sum_{s,k=0}^{d-1} A_s(k)^{2\alpha}$ and 
$A_s(k)=\sum_{x=0}^{d-1}(-1)^{k\cdot x}\sqrt{\lambda_x\lambda_{x\oplus s}}$. 
Here $x$, $s$, and $k$ denote $m$-bit binary strings, $\oplus$ is the
bitwise XOR operation, and
$k\cdot x=\sum_{j=0}^{m-1}k_jx_j \pmod 2$
is the binary inner product~\cite{Gianpaolo2026,Fabio2026}.
This construction eliminates the explicit local-unitary optimization
and expresses the nonlocal SRE directly in terms of the entanglement
spectrum. 
Its validity is supported by exact results for low-rank spectra and
extensive numerical optimization \cite{Gianpaolo2026,Fabio2026,Liu2026}. 
The reference state has already been shown to be a stationary point of the
local-unitary optimization~\cite{Liu2026}. 
Here we provide further analytical evidence for this construction.
For the reference state associated with the ordered entanglement spectrum,
we found that the local-unitary optimization can be reduced from two
independent unitaries to a single one~\cite{SM}, 
\begin{equation}
\min_{U_A,U_B\in U(d)}
M_{\alpha}\!\left(U_A\otimes U_B\ket{\tilde{\psi}}\right)
=
\min_{U\in U(d)}
M_{\alpha}\!\left(U\otimes U^{*}\ket{\tilde{\psi}}\right).
\end{equation}
We then consider an arbitrary infinitesimal deformation
$U(t)=e^{-iKt}$, where $K$ is any Hermitian generator and $t\in\mathbb{R}$.
For integer $\alpha\geq2$, we prove that~\cite{SM}
\begin{equation}
\left.
\frac{\partial^2}{\partial t^2}
M_{\alpha}\!\left[
U(t)\otimes U^{*}(t)\ket{\tilde{\psi}}
\right]
\right|_{t=0}
\geq 0 .
\end{equation}
Together with the previously established stationarity of the reference
state~\cite{Liu2026,SM}, this shows that it is a local minimum of the SRE along arbitrary
local-unitary directions.

The introduced spectral formulation naturally raises the question of which features of the ES control the SRE measure of nonlocal nonstabilizerness. 
A natural candidate is the anti-flatness $\mathcal{A}=\sum_i\lambda_i^3-(\sum_i\lambda_i^2)^2$, which quantifies deviations from spectral uniformity and has previously been connected to nonstabilizerness~\cite{Tirrito2024}. 
To extend this connection, we introduce the generalized anti-flatness 
\begin{equation}
    \mathcal{A}_{\alpha}= P_{2\alpha-1}-P_{\alpha}^{2} \nonumber=\sum_{i<j}\lambda_i\lambda_j
    \left(
    \lambda_i^{\alpha-1}-\lambda_j^{\alpha-1}
    \right)^2,
\end{equation}
with $P_{\alpha}=\sum_i\lambda_i^{\alpha}$. This generalized anti-flatness reduces to the conventional anti-flatness when $\alpha=2$. 

After introducing the nonlocal SRE $M_{\alpha}^{\mathrm{NL}}$ and the generalized anti-flatness $\mathcal{A}_{\alpha}$, we define the $\alpha$-R\'enyi entanglement entropy as
$S_{\alpha}=(1-\alpha)^{-1}\log_2 P_{\alpha}$. Our main result is the pair of spectral bounds~\cite{SM}
\begin{equation}
    \frac{1}{\alpha-1}
    \log_{2}\frac{1}{1-2\alpha\mathcal{A}_{\alpha}}
    \le
    M_{\alpha}^{\mathrm{NL}}
    \le
    2S_{\alpha}.
\end{equation}
The upper bound holds for arbitrary $\alpha>0$ with $\alpha\neq1$, whereas the lower bound applies to integer R\'enyi indices $\alpha\geq2$. The upper bound for $\alpha=2$ was previously established in Refs.~\cite{Cao2025,Fabio2026,Gianpaolo2026}, and here we extend it to general $\alpha$. Using $\ln x\leq x-1$, the lower bound further implies
$M_{\alpha}^{\mathrm{NL}}\ge [2\alpha/((\alpha-1)\ln 2)]\mathcal{A}_{\alpha}>\mathcal{A}_{\alpha}$,
showing that the generalized anti-flatness by itself provides a direct lower bound on the nonlocal SRE.

These bounds reveal two complementary spectral controls of nonlocal nonstabilizerness. The R\'enyi entropy constrains the overall amount supported by bipartite entanglement, while the generalized anti-flatness quantifies the minimum contribution enforced by spectral non-uniformity. Together, they provide a direct link between universal entanglement-spectrum structure and nonlocal nonstabilizerness.

\textit{\textbf{Simple decaying spectrum}.---}
To gain physical insight into the spectral bounds derived above, we consider two representative classes of entanglement spectra with qualitatively different decay behaviors. For simplicity, we restrict ourselves to Schmidt ranks of the form $\chi=2^m$ with $m$ a positive integer, and denote $N\equiv2^m$ in the following. The first is an exponentially decaying spectrum, $\lambda_x=(1-q)q^x/(1-q^N)$ for $x=0,\ldots,N-1$ and $0<q<1$, which interpolates between a strongly nonuniform spectrum at $q\rightarrow0$ and an almost flat spectrum at $q\rightarrow1$. The second is an algebraically decaying spectrum, $\lambda_r=r^{-p}/H(N,p)$ for $r=1,\ldots,N$, where $H(N,p)=\sum_{r=1}^N r^{-p}$ and $p>0$ controls the spectral tail. These two examples provide simple settings in which the roles of EE and spectral non-uniformity can be contrasted directly.

For the exponentially decaying spectrum, all spectral moments can be obtained analytically, which allows both the R\'enyi entropy and the anti-flatness to be evaluated exactly. More importantly, the nonlocal SRE can also be determined analytically. In the weakly entangled limit $q\to0$, we find $M_2^{\rm NL}\simeq 4q/\ln2$, $\mathcal{A}_2\simeq q$, and $S_2\simeq2q/\ln2$, leading to
\begin{equation}
    M_2^{\rm NL}
    \simeq
    \frac{4}{\ln2}\mathcal{A}_2
    \simeq
    2S_2 .
\end{equation}
Thus, both the anti-flatness lower bound and the entanglement-entropy upper bound become asymptotically tight. This shows that close to the product-state limit, entanglement, spectral non-uniformity, and nonlocal nonstabilizerness are governed by the same leading spectral deformation.
The opposite limit $q\to 1$ reveals a rather different behavior. The spectrum becomes nearly flat and $S_2$ approaches its maximal value $\log_2N$, while both $\mathcal{A}_2$ and $M_2^{\rm NL}$ vanish. To leading order, we have 
\begin{equation}
    M_2^{\rm NL}\simeq \frac{N^2}{\ln2}\mathcal{A}_2\simeq \log_2N-S_2
\end{equation}
Hence, a highly entangled state need not possess large nonlocal nonstabilizerness. What matters is not only the total amount of entanglement, but also how strongly the ES deviates from flatness.

For an algebraically decaying spectrum, $\lambda_r=r^{-p}/H(N,p)$, the large-$N$ behavior is summarized in Table~\ref{spec_alg}. For $p>1$, both $S_2$ and $\mathcal{A}_2$ remain finite, and correspondingly $M_2^{\mathrm{NL}}=O(1)$. At the marginal point $p=1$, the entropy grows as $S_2\sim2\log_2\ln N$, while the anti-flatness instead vanishes as $\mathcal{A}_2\sim\zeta(3)/(\ln N)^3$ with $\zeta(\cdot)$ the Riemann zeta function.
For $p\in(0,1)$, this separation becomes even more pronounced. The R\'enyi entropy grows as $O(\log N)$, whereas $\mathcal{A}_2\to0$ and the nonlocal SRE remains bounded, $M_2^{\mathrm{NL}}=O(1)$. Thus, algebraically broad spectra provide a clear example in which increasing entanglement does not necessarily imply increasing nonlocal nonstabilizerness, emphasizing the importance of the detailed structure of the ES.

\begin{table}[b]
    \caption{
    \label{spec_alg}
    Asymptotic behavior of $S_2$, $\mathcal{A}_2$, and $M_2^{\mathrm{NL}}$ for algebraically decaying entanglement spectra. While all quantities remain finite for $p>1$, a qualitative separation emerges for $p\leq1$, where the EE grows with $N$ but the anti-flatness vanishes. The nonlocal SRE, however, remains $O(1)$ throughout $0<p<1$ and grows only as $O(\log_2\ln N)$ at the marginal point $p=1$, illustrating that increasing entanglement does not necessarily imply increasing nonlocal nonstabilizerness. 
    } 
    \begin{ruledtabular}
    \begin{tabular}{cccc}
    $p$            & $S_2$                      & $\mathcal{A}_2$        & $M_2^{\mathrm{NL}}$ \\ 
    \hline
    $(1, +\infty)$ & $O(1)$                     & $O(1)$               & $O(1)$                           \\
    $1$            & $2\log_2\ln N + O(1)$      & $\zeta(3)/(\ln N)^3$ & $O(\log_2\ln N)$                 \\
    $(1/2, 1)$     & $2(1-p)\log_2 N + O(1)$    & $N^{-3(1-p)}$        & $O(1)$                           \\
    $1/2$          & $\log_2N-\log_2\ln N+O(1)$ & $N^{-3/2}$           & $O(1)$                           \\
    $(1/3, 1/2)$   & $\log_2 N+O(1)$            & $N^{-3(1-p)}$        & $O(1)$                           \\
    $1/3$          & $\log_2 N+O(1)$            & $N^{-2}\ln N$        & $O(1)$                           \\
    $(0, 1/3)$     & $\log_2 N+O(1)$            & $N^{-2}$             & $O(1)$                           \\
    \end{tabular}
    \end{ruledtabular}
\end{table}

These two models therefore demonstrate that nonlocal nonstabilizerness is sensitive to the detailed structure of the ES rather than to EE alone. Rapidly decaying spectra in the weak-entanglement regime exhibit a nearly one-to-one relation between entropy, anti-flatness, and nonlocal SRE, whereas broad algebraic spectra allow entanglement to grow parametrically without a corresponding growth of nonlocal nonstabilizerness. This separation highlights the complementary roles of entropy and anti-flatness in characterizing the nonstabilizer content encoded in the ES. 

\begin{figure*}
    \includegraphics[width=\linewidth]{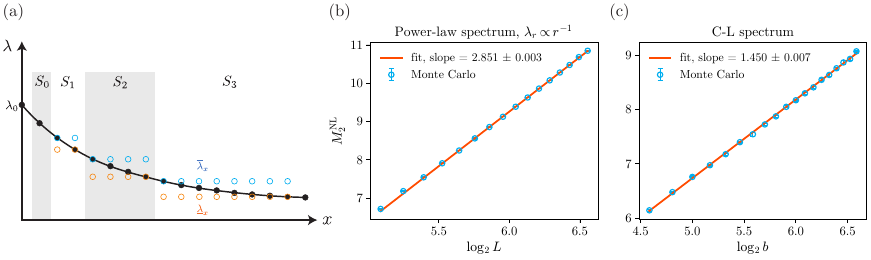}
    \caption{
    (a) Schematic illustration of the shell-based sandwich construction. Within each dyadic shell $S_j$, the largest and smallest eigenvalues are used to construct upper and lower shell-flat spectra, respectively, which bound the original ordered ES from above and below. (b) Monte Carlo results for the algebraically decaying spectrum at $p=1$, with $L=\log_2N$. The fitted slope is $2.851\pm0.003$, approaching the asymptotic value $3$. (c) Monte Carlo results for the C--L spectrum, with fitted slope $1.450\pm0.007$, approaching the predicted value $3/2$.
    \label{fig1}
    }
\end{figure*}

\textit{\textbf{Calabrese--Lefevre spectrum}.---}
We finally turn to the Calabrese--Lefevre (C--L) spectrum, which describes the universal distribution of Schmidt eigenvalues in one-dimensional critical systems~\cite{Calabrese2008}. Denoting the largest eigenvalue by $\lambda_{\max}$ and $b=-\ln\lambda_{\max}$, the mean number of eigenvalues larger than a given
$\lambda$ takes the form $n(\lambda)=I_0\left(2\sqrt{b\ln(\lambda_{\max}/\lambda)}\right)$. In a conformal critical system, $b=(c/6)\ln L+O(1)$, where $c$ is the central charge and $L$ is the subsystem size. The C--L spectrum therefore provides a direct route for translating universal critical entanglement structure into the scaling of nonlocal nonstabilizerness. 

To extract the nonlocal SRE, we organize the ordered Schmidt spectrum into dyadic shells
$S_j=\{2^j,\ldots,2^{j+1}-1\}$ and denote the weight of each shell by
$\mu_j=\sum_{x\in S_j}\lambda_x$.
For spectra that are flat within each shell, we prove that
$\zeta_\alpha=2^{(1-\alpha)M_\alpha^{\rm NL}}$ is bounded solely in terms of the shell weights as
\begin{equation}
    \lambda_0^{2\alpha}+\sum_j \mu_j^{2\alpha}
    \leq
    \zeta_\alpha
    \leq
    CZ\delta^{2\alpha-1},
\end{equation}
where $\delta=\max_j\{\mu_j\}$, $Z$ is the total weight of the spectrum and $C$ is an independent constant.
We note this bound does not require the spectrum to be normalized.
It shows that the asymptotic scaling of the nonlocal SRE is governed by the largest shell weight together with the number of shells carrying comparable weight. 

The actual C--L spectrum is not exactly flat within each shell. 
We therefore construct upper and lower shell-flat spectra, $\overline{\lambda}$ and $\underline{\lambda}$, satisfying
$\underline{\lambda}_x\leq\lambda_x\leq\overline{\lambda}_x$ for all $x$, as illustrated in Fig.~\ref{fig1}(a).
We refer to this construction as the sandwich method. 
Defining 
$\mu_j^+=2^j\lambda_{2^j}=\sum_{x\in S_j}\overline{\lambda}_x$
and
$\mu_j^-=2^j\lambda_{2^{j+1}-1}=\sum_{x\in S_j}\underline{\lambda}_x$,
together with
$\delta_+=\max\{\lambda_0,\sup_j\mu_j^+\}$
and
$Z_+=\lambda_0+\sum_j\mu_j^+$,
the ordering of the spectrum immediately yields the squeezing bound
\begin{equation}
    \lambda_0^4+\sum_j(\mu_j^-)^4
    \leq
    \zeta_2
    \leq
    C Z_+\delta_+^3,
\end{equation}
where we consider $\alpha=2$. 
Moreover, $Z_+=O(1)$, so the asymptotic behavior is entirely determined by the scaling of $\delta_+$ and the lower shell weights $\mu_j^-$. 

The large-$b$ asymptotics of the C--L spectrum give $\delta_+=O(b^{-1/2})$. At the same time, there exists a window of shell indices with width $|\mathcal{J}_A(b)|=\Theta(\sqrt{b})$ for which $\mu_j^-\gtrsim b^{-1/2}$. Hence the lower bound contains $\Theta(\sqrt{b})$ terms of order $b^{-2}$, while the upper bound scales as $b^{-3/2}$. The two bounds therefore match asymptotically, which gives  
$\zeta_2\sim b^{-3/2}$. 
Consequently, from $M_2^{\rm NL}=-\log_2\zeta_{2}$ we obtain 
\begin{equation}
    M_2^{\rm NL}
    \sim
    \frac{3}{2}\log_2 b\sim
    \frac{3}{2}\log_2\ln L+\frac{3}{2}\log_2\frac{c}{6}+O(1). 
\end{equation}
Thus, at a one-dimensional conformal critical point, the nonlocal SRE exhibits a universal double-logarithmic growth with subsystem size, as first conjectured in Ref.~\cite{Gianpaolo2026}. Importantly, the leading coefficient $3/2$ is independent of the central charge, which enters only through the additive $O(1)$ contribution via $b=(c/6)\ln L+O(1)$. 

The same shell argument extends to integer R\'enyi index $\alpha\geq2$. In this case the corresponding spectral quantity obeys $\zeta_\alpha=\Theta\left(b^{-\alpha+1/2}\right)$, leading to
\begin{equation}
    M_\alpha^{\rm NL}
    \sim
    \frac{2\alpha-1}{2\alpha-2}
    \log_2\ln L+O(1)
\end{equation}
which is very different from the scaling behaviour of R\'enyi entanglemnt entropy $S_{\alpha}\sim(1+1/\alpha)c/6\ln L$~\cite{Pasquale2004,Pasquale2009}. 
The C--L spectrum therefore predicts a universal family of double logarithmic scalings for nonlocal nonstabilizerness at one-dimensional conformal critical points, with a R\'enyi-index-dependent coefficient but no dependence on the central charge at leading order. 

The sandwich method developed above can also sharpen the result for the algebraically decaying spectrum discussed earlier. At the marginal point $p=1$, the spectrum is $\lambda_r=[rH(N,1)]^{-1}$ with $H(N,1)\sim\ln N$. The weight carried by each dyadic shell is therefore of order $(\ln N)^{-1}$, while the number of contributing shells grows as $O(\ln N)$. Applying the shell bounds then yields $\zeta_{\alpha}=\Theta[(\ln N)^{-(2\alpha-1)}]$, and hence 
\begin{align}
    M_{\alpha}^{\rm NL}\sim \frac{2\alpha-1}{\alpha-1}\log_2\ln N \sim \frac{2\alpha-1}{\alpha-1}\log_2 L.
\end{align}
Thus, the sandwich method not only reproduces the upper bound obtained previously, but also fixes the asymptotic scaling exactly at the marginal algebraic decay $p=1$. 

We numerically test these predictions for both the marginal algebraic spectrum and the critical Ising ES using Monte Carlo sampling~\cite{SM,Sierant2026}, since accessing the asymptotic regime requires very large Schmidt ranks. For the Ising case, the critical ES is constructed independently from the corner-transfer-matrix  solution~\cite{Peschel1999,Davies1988,Ingo2009}, rather than directly from the C--L distribution, thereby providing an independent test of the predicted C--L scaling.
In both cases, the numerically extracted coefficients are slightly smaller than the theoretical predictions as shown in Fig.~\ref{fig1}(b,c). We attribute this discrepancy mainly to finite-size effects, since the analytical coefficients describe the asymptotic large-size regime, whereas the numerically accessible spectra have not yet fully reached this limit. The effective coefficients therefore approach the asymptotic values from below, and the remaining deviation is expected to decrease as larger Schmidt ranks become accessible. Despite these pre-asymptotic corrections, the numerical results are consistent with the predicted logarithmic scaling forms and clearly distinguish the two scaling coefficients.

\textit{\textbf{Discussion and outlook}. ---}
Our results establish a unified spectral perspective on bipartite
nonlocal nonstabilizerness. The R\'enyi entropy constrains the overall
magnitude of the nonlocal SRE, while the generalized anti-flatness
captures the minimum contribution enforced by spectral non-uniformity.
We further show that, for integer $\alpha\geq2$, the ordered Schmidt
reference state is a local minimum of the SRE under local-unitary
variations, providing additional analytical support for the
Schmidt-spectrum construction. The exponentially and algebraically 
decaying spectra show that entanglement and nonlocal nonstabilizerness
can either track each other closely or become parametrically separated,
demonstrating that the latter depends on more detailed spectral
information than EE alone. This distinction becomes particularly clear
for the C--L spectrum, where the R\'enyi EE scales as
$S_\alpha\sim\ln L$, whereas the nonlocal SRE grows only as
$M_\alpha^{\mathrm{NL}}\sim[(2\alpha-1)/(2\alpha-2)]\log_2\ln L$
for integer $\alpha\geq2$. The leading coefficient is independent of
the central charge, revealing a universal but parametrically weaker
growth of nonlocal nonstabilizerness at one-dimensional conformal
critical points. Our Monte Carlo results for both the marginal
algebraic spectrum and the critical Ising spectrum are consistent with
these asymptotic predictions.

Several open directions follow naturally from this work. A central theoretical problem is to prove whether the Schmidt reference state globally minimizes the SRE under arbitrary local-unitary transformations, which would promote the present spectral construction to an exact characterization of nonlocal nonstabilizerness. It would also be interesting to extend the shell-based analysis to non-integer R\'enyi indices and to determine whether similar universal scaling structures survive in higher-dimensional critical systems, non-conformal transitions, and non-equilibrium dynamics~\cite{Zhang2026,Tirrito2025transport,Tirrito2025,Ebner2026}. Another important direction is to understand how much of the nonlocal SRE can be inferred from a finite set of spectral moments, which may lead to experimentally accessible estimators or bounds~\cite{Liu2026}. More broadly, the present framework suggests that nonlocal nonstabilizerness can serve as a complementary probe of entanglement-spectrum organization, capable of distinguishing states with comparable entanglement entropy but qualitatively different internal spectral structures.

\textit{\textbf{Acknowledgments}.---}
This work is supported by the National Key Research and Development 
of China (Grant No.~2021YFA1402001) and the National Natural Science Foundation of China (NSFC) (Grant No.~12375007). 

\textit{Note added.---}
While completing this manuscript, we became aware of Ref.~\cite{Sierant202608},
which gives an exact spectral characterization of nonlocal magic based on
the stabilizer fidelity. That work also highlights the role of dyadic rank
sectors and logarithmic-scale Schmidt-weight distributions, which
conceptually overlap with the dyadic-shell framework developed here.
The two works nevertheless focus on different nonstabilizerness measures
and use different analytical approaches. The results in the overlapping
regimes are fully consistent.

\nocite{*}
\bibliography{ref}

\clearpage
\appendix
\onecolumngrid

\section{General expression of nonlocal SRE}
The general $\alpha$-SRE is defined as~\cite{Lorenzo2022}
\begin{align}
    M_{\alpha}(\ket{\psi}) := \frac{1}{1-\alpha}\left[\log_2\left(
        \sum_{P\in\mathcal{P}_n}\braket{\psi|P|\psi}^{2\alpha}\right)
    -\log_2d\right]
    =\frac{1}{1-\alpha}\log_2 \zeta_{\alpha}(\ket{\psi}), 
\end{align}
where $d=2^n$ the Hilbert space dimension, $n$ is the qubit number of state $\ket{\psi}$, $\mathcal{P}_n$ denotes the set of all $n$-qubit Pauli strings
with positive phase and 
\begin{align}
    \zeta_{\alpha}(\ket{\psi})=\frac{1}{d}\sum_{P\in\mathcal{P}_n}\braket{\psi|P|\psi}^{2\alpha}.
\end{align} 
According to Refs.~\cite{Fabio2026,Gianpaolo2026,Liu2026}, it is conjectured that the Schmidt reference state 
\begin{align}
    \ket{\psi_{\mathrm{Sch}}}=\sum_{x\in\mathbb{F}_2^m}\sqrt{\lambda_x}\ket{x},
\end{align}
reaches the minimum SRE for all $\alpha$, therefore we write  
\begin{align}
    M_{\alpha}^{\mathrm{NL}}=M_{\alpha}^{\mathrm{Sch}} := M_{\alpha}(\ket{\psi_{\mathrm{Sch}}}).
\end{align}
for any given sorted entanglement spectrum $\{\lambda_i\}_{i=0}^{2^m-1}$ with $\lambda_0\ge\lambda_1\ge ...\ge\lambda_{2^m-1}$. 

Denoting Pauli string $W\in\mathcal{P}_m$ in the form of  
\begin{align}
    W_{s,k}=i^{s\cdot k}X^s Z^k,\quad s, k\in\mathbb{F}_2^m
\end{align}
so we have 
\begin{align}
    \braket{\psi_{\mathrm{Sch}}|W_{s,k}|\psi_{\mathrm{Sch}}}= 
    i^{s\cdot k}\sum_{x,y}\sqrt{\lambda_x\lambda_y}(-1)^{k\cdot x}\braket{y|x\oplus s}
    =i^{s\cdot k}\sum_x \sqrt{\lambda_x\lambda_{x\oplus s}}(-1)^{k\cdot x},
\end{align}
and we denote 
\begin{align}
    A_s(k): = \sum_x \sqrt{\lambda_x\lambda_{x\oplus s}}(-1)^{k\cdot x},
\end{align}
then we can write  
\begin{align}
    \zeta_{\alpha}(\ket{\psi_{\mathrm{Sch}}}) = \frac{1}{d}\sum_{s,k=0}^{2^m-1} A_s(k)^{2\alpha}
\end{align}
therefore we obtain expression for general $\alpha$-nonlocal SRE: 
\begin{align}
    M_{\alpha}^{\mathrm{Sch}} := M_{\alpha}(\ket{\psi_{\mathrm{Sch}}})=
    \frac{1}{1-\alpha}\log_2\left(\frac{1}{d}\sum_{s,k=0}^{2^m-1} A_s(k)^{2\alpha}\right).
\end{align}
We note that $\braket{\psi_{\mathrm{Sch}}|W_{s,k}|\psi_{\mathrm{Sch}}}$ must be real because any Pauli string is Hermitian, 
so we can also write 
\begin{align}
    \zeta_{\alpha}^{\mathrm{Sch}} := \zeta_{\alpha}(\ket{\psi_{\mathrm{Sch}}}) = 
    \frac{1}{d}\sum_{s,k=0}^{2^m-1} |A_s(k)|^{2\alpha}.
\end{align}
This may avoid misunderstaing when $\alpha=1/2,3/2,5/2,...$. 

\section{Single-unitary reduction}
Suppose we consider the bipartite state $\ket{X}_{AB}$
\begin{align}
    \ket{X}_{AB}=\sum_{i,j\in\mathbb{F}_2^m}X_{ij}\ket{i}_A\ket{j}_B,
\end{align}
then we can write 
\begin{align}
    \zeta_{\alpha}(\ket{X})=\frac{1}{d^2}\sum_{P,Q\in\mathcal{P}_m}
    \left[\operatorname{Tr}\left(X^{\dagger}PXQ^{\top}\right)\right]^{2\alpha},
\end{align}
where $d=2^m$ and $n=2m$ is the total qubit number. In general we have 
\begin{align}
    M_{\alpha}(\ket{X})=\frac{1}{1-\alpha}\log_2\left(\frac{1}{2^{2m}} \sum_{P,Q\in\mathcal{P}_m}
    \left[\operatorname{Tr}\left(X^{\dagger}PXQ^{\top}\right)\right]^{2\alpha}\right),
\end{align}
so when $\alpha\ge 2$, the minimization of $M_{\alpha}(\ket{X})$ equals to the maximization 
of $\zeta_{\alpha}(\ket{X})$. 

Denoting 
\begin{align}
    \Omega = \frac{1}{2^m}\sum_{P\in\mathcal{P}_m}P^{\otimes 2\alpha},
\end{align}
we can write 
\begin{align}
    \zeta_{\alpha}(\ket{X}) = \operatorname{Tr}\left[
        \left(X^{\dagger}\right)^{\otimes 2\alpha}\Omega X^{2\alpha}\Omega
    \right],\label{zetaX}
\end{align}
where we use $\Omega=\Omega^{\top}$. We now do singular value decomposition for matrix $X$, 
\begin{align}
    X = UDV^{\dagger},
\end{align}
where $U$ and $V^{\dagger}$ are unitaries and $D$ is a non-negative diagonal matrix. In fact 
these diagonal elements are the entanglement spectrum 
\begin{align}
    D = \operatorname{diag}(\sqrt{\lambda_0},\sqrt{\lambda_1},...). 
\end{align}
So by replacing $X$ in Eq.~\eqref{zetaX}, we obtain, 
\begin{align}
    \zeta_{\alpha}(\ket{X}) = \zeta_{\alpha}(UDV^{\dagger})&=\operatorname{Tr}\left[
        \left(VDU^{\dagger}\right)^{\otimes 2\alpha} \Omega 
        \left(UDV^{\dagger}\right)^{\otimes 2\alpha} \Omega
    \right]\\
    &= \operatorname{Tr}\left[
        D^{\otimes 2\alpha} \left(U^{\dagger}\right)^{\otimes 2\alpha}\Omega 
        U^{\otimes 2\alpha} D^{\otimes 2\alpha} \left(V^{\dagger}\right)^{\otimes 2\alpha}\Omega V^{\otimes 2\alpha}
    \right],
\end{align}
and if we define 
\begin{align}
    K_U=\left(\sqrt{D}\right)^{\otimes 2\alpha}\left(U^{\dagger}\right)^{\otimes 2\alpha} \Omega U^{\otimes 2\alpha}\left(\sqrt{D}\right)^{\otimes 2\alpha},\\
    K_V=\left(\sqrt{D}\right)^{\otimes 2\alpha}\left(V^{\dagger}\right)^{\otimes 2\alpha} \Omega V^{\otimes 2\alpha}\left(\sqrt{D}\right)^{\otimes 2\alpha},
\end{align}
we note that $K_U$ and $K_V$ are hermitian, 
then we can write 
\begin{align}
    \zeta_{\alpha}(UDV^{\dagger}) = \operatorname{Tr}\left(K_UK_V\right), 
\end{align}
so by using Hilbert-Schmidt Cauchy-Schwarz inequality we have 
\begin{align}
    \zeta_{\alpha}(UDV^{\dagger})=\operatorname{Tr}\left(K_UK_V\right)\le 
    \sqrt{\operatorname{Tr}\left(K_UK_U\right)\operatorname{Tr}\left(K_VK_V\right)} = \sqrt{\zeta_{\alpha}(UDU^{\dagger})\zeta_{\alpha}(VDV^{\dagger})}, 
\end{align}
and everything is positive here. This inequality means, 
\begin{align}
    \zeta_{\alpha}(UDV^{\dagger})\le \max\{\zeta_{\alpha}(UDU^{\dagger}),\zeta_{\alpha}(VDV^{\dagger}) \}, 
\end{align}
therefore, the optimization can only be restricted within 
\begin{align}
    \max_{U,V\in U(2^m)} \zeta_{\alpha}(UDV^{\dagger})=\max_{U\in U(2^m)}\zeta_{\alpha}(UDU^{\dagger}), 
\end{align}
so for SRE, we only need to optimize
\begin{align}
    \min_{U_A,U_B\in U(2^m)} M_{\alpha}\left(U_A\otimes U_B\ket{\psi_{\mathrm{Sch}}}_{AB}\right)=
    \min_{U\in U(2^m)} M_{\alpha}\left(
        U\otimes U^{*}\ket{\psi_{\mathrm{Sch}}}
    \right).
\end{align}
This helps us reduce much of the optimization resources. 

\section{Local minimality of the Schmidt reference state}
Now we can only consider $H=U\Sigma U^{\dagger}$ where 
\begin{align}
    \Sigma = \operatorname{diag}\left(\sqrt{\lambda_0},\sqrt{\lambda_1},...\right). 
\end{align}
For integer $\alpha\ge 2$, we show in the following that 
\begin{align}
    \frac{\mathrm{d}}{\mathrm{d} t}\zeta_{\alpha}(e^{iKt}\Sigma e^{-iKt})\Bigg|_{t=0} &= 0,\\
    \frac{\mathrm{d}^2}{\mathrm{d} t^2}\zeta_{\alpha}(e^{iKt}\Sigma e^{-iKt})\Bigg|_{t=0} &\le 0,
\end{align}
where $K$ is any hermitian operator, $t\in\mathbb{R}$ and we denote $H(t)=e^{iKt}\Sigma e^{-iKt}$ in the follwoing. 
Here we do require the elements in $\Sigma$ are arranged in descending order.

We first introduce the follwoing lemma. 
For density matrix $\rho\in\mathbb{C}^{d\times d}$, $\rho \ge 0$ and positive integer $r$, the matrix 
\begin{align}
    \mathcal{T}_r(\rho) = \frac{1}{d}\sum_{P\in\mathcal{P}_m}\left[\operatorname{Tr}(\rho P)\right]^r P
\end{align}
is semi-positive, and its trace $\operatorname{Tr}\left(\mathcal{T}_r(\rho)\right)=\left[\operatorname{Tr}(\rho)\right]^r$, 
where $d=2^m$ here. 

\begin{proof}
    We can write 
    \begin{align}
        \left[\operatorname{Tr}(\rho P)\right]^r = \operatorname{Tr}\left(\rho^{\otimes r} P^{\otimes r}\right).
    \end{align}
    We further denote $\overline{X}=X^{\otimes r}$ and $\overline{Z}=Z^{\otimes r}$, and they can be regarded as logical qubit Pauli matrices, 
    \begin{align}
        \overline{X}^2=\overline{Z}^2=\mathbb{I},\quad \overline{X}\overline{Z}=-\overline{Z}\overline{X}. 
    \end{align}
    So for a logical qubit $z$ and $r-1$ auxiliary qubits $w_1,w_2,...,w_{r-1}$, we define the operator $V$ as 
    \begin{align}
        V\ket{z,w_1,w_2,...,w_{r-1}}=\ket{z\oplus w_1,z\oplus w_2,...,z\oplus w_r},
    \end{align}
    where $w_r=w_1\oplus w_2\oplus...\oplus w_{r-1}$. The operator $V$ is a kind of permutation, so it is unitary. 
    Now we have 
    \begin{align}
        V(X\otimes \mathbb{I}^{\otimes r-1})V^{\dagger} = X^{\otimes r}, \quad 
        V(Z\otimes \mathbb{I}^{\otimes r-1})V^{\dagger} = Z^{\otimes r},
    \end{align}
    this is because 
    \begin{align}
        V(X\otimes \mathbb{I}^{\otimes r-1})\ket{z,w_1,w_2,...,w_{r-1}}&=V\ket{z\oplus 1,w_1,w_2,...,w_{r-1}}\\
        &= \ket{z\oplus 1\oplus w_1,z\oplus 1\oplus w_2,..., z\oplus 1\oplus w_r}\\
        &= X^{\otimes r} \ket{z\oplus w_1,z\oplus w_2,...,z\oplus w_r}=X^{\otimes r} V \ket{z,w_1,w_2,...,w_{r-1}},
    \end{align}
    and similarly for $Z$, 
    \begin{align}
        V(Z\otimes \mathbb{I}^{\otimes r-1})\ket{z,w_1,w_2,...,w_{r-1}} & = (-1)^z\ket{z\oplus w_1,z\oplus w_2,...,z\oplus w_r}, \\
        Z^{\otimes r} V \ket{z,w_1,w_2,...,w_{r-1}} &= (-1)^{z}\ket{z\oplus w_1,z\oplus w_2,...,z\oplus w_r}. 
    \end{align}
    Here we note that $r$ is odd. 
    For $Y$ we have 
    \begin{align}
        V(Y\otimes \mathbb{I}^{\otimes r-1})V^{\dagger} = i X^{\otimes r} Z^{\otimes r}=i^{1-r} Y^{\otimes r} = (-1)^{(r-1)/2}Y^{\otimes r}. 
    \end{align}
    Define partial trace on anxiliary space 
    \begin{align}
        \tau = \operatorname{Tr}_{\mathrm{aux}}\left(V^{\dagger}\rho^{\otimes r V}\right),
    \end{align}
    because all operations above are semi-definite preservative, so $\tau\ge 0$. 
    And its Pauli coefficients are 
    \begin{align}
        \operatorname{Tr}(\tau P) &= \operatorname{Tr}[V^{\dagger}\rho^{\otimes r}V(P\otimes \mathbb{I}^{\otimes r-1})]\\
        &= \operatorname{Tr}[\rho^{\otimes r}V(P\otimes \mathbb{I}^{\otimes r-1})V^{\dagger}]\\
        &= \eta_r(P)\operatorname{Tr}(\rho^{\otimes r} P^{\otimes r})=\eta_r(P) [\operatorname{Tr}(\rho P)]^r,
    \end{align}
    so we can write 
    \begin{align}
        \tau = \frac{1}{d}\sum_{P\in\mathcal{P}_m}\eta_r(P)[\operatorname{Tr}(\rho P)]^r P.
    \end{align}
    We now deal with the sign $\eta_r(P)$, 
    \begin{align}
        \eta_r(P) = (-1)^{(r-1)/2 * n_Y(P)}
    \end{align}
    where $n_Y(P)$ is the number of $Y$ in string $P$. 
    When $(r-1)/2$ is even, i.e., $r=1 \mod 4$, 
    \begin{align}
        \mathcal{T}_r(\rho)=\tau\ge 0,
    \end{align}
    when $(r-1)/2$ is odd, i.e., $r=3\mod 4$, 
    \begin{align}
        \eta_r(P)=(-1)^{n_Y(P)}, 
    \end{align}
    on the other side, $P^{\top}=(-1)^{n_Y(P)}P$, 
    so we have 
    \begin{align}
        \tau^{\top}=\frac1d\sum_{P\in\mathcal{P}_m} \eta_r(P)[\operatorname{Tr}(\rho P)]^r P^{\top} = \mathcal{T}_r(\rho)
    \end{align}
    and $\tau\ge 0$ so $\tau^{\top}\ge 0$, therefore $\mathcal{T}_r(\rho)\ge 0$. 
\end{proof}

We now calculate $\zeta_{\alpha}(\Sigma)$, and we denote 
\begin{align}
    \zeta_{\alpha}(\Sigma) = \frac{1}{d}\sum_{P\in\mathcal{P}_m}\braket{f|P|f}^{2\alpha}
\end{align}
where $\ket{f}=(f_0,f_1,...)^{\top}$ with $f_i=\sqrt{\lambda_i}$. 
We first calculate the second order derivative of $\zeta_{\alpha}(f):=\zeta_{\alpha}(\Sigma)$, 
\begin{align}
    \zeta_{\alpha}(f+th) = \frac{1}{d}\sum_{P\in\mathcal{P}_m}
    \left[\braket{f|P|f}+2\operatorname{Re}\left(\braket{f|P|h}\right)t + \braket{h|P|h}t^2\right]^{2\alpha}, 
\end{align}
so we obtain 
\begin{align}
    \frac{\partial}{\partial t} \zeta_{\alpha}(f+th) &= \frac{1}{d}\sum_{P\in\mathcal{P}_m}\left[
        2\alpha(f_P + a_P t + h_P t^2)^{2\alpha-1}(a_P+2h_P t)
    \right],\\
    \frac{\partial^2}{\partial t^2} \zeta_{\alpha}(f+th)&=\frac{1}{d}\sum_{P\in\mathcal{P}_m}\left[
    2\alpha(f_P+a_P t+h_P t^2)^{2\alpha-2}\left[(2\alpha-1)a_P^2+(8\alpha-2)h_P^2 t^2+(8\alpha-2)h_P a_P t+2h_Pf_P\right]
    \right],
\end{align}
where we use $f_P=\braket{f|P|f}$, $a_P=2\operatorname{Re}(\braket{f|P|h})$ and $h_P=\braket{h|P|h}$. 
Clearly, when $t\to 0$, we have 
\begin{align}
    \frac{\partial^2}{\partial t^2} \zeta_{\alpha}(f+th)\Bigg|_{t=0}&=\frac1d\sum_{P\in\mathcal{P}_m}
    2\alpha(2\alpha-1)f_P^{2\alpha-2}a_P^2+4\alpha f_{P}^{2\alpha-1} h_P\\
    &= \frac{2\alpha(2\alpha-1)}{d}\sum_{P\in\mathcal{P}_m}f_P^{2\alpha-2}a_P^2 + 4\alpha\braket{h|\mathcal{T}_{2\alpha-1}(\ket{f}\bra{f})|h}\ge 0. 
\end{align}
This is because $2\alpha-2$ and $2$ are even in the first term, and the second term by using abovementioned lemma, is also 
always non-negative. 
Therefore, $\zeta_{\alpha}(f)$ is convex along any direction $h$ for any $f$. 
If we set $h\neq 0$, and denote $\phi(t)=\zeta_{\alpha}(f+th)$, 
then it is a polynomial of $t$, explicitly, we can write 
\begin{align}
    \phi(t) = \frac{1}{d}\sum_{P\in\mathcal{P}_m}\left(f_P+a_P t +h_P t^2\right)^{2\alpha}=\frac{1}{d}\sum_{P\in\mathcal{P}_m}\left(h_P^{2\alpha}t^{4\alpha}+ 2\alpha a_P h_P^{2\alpha-1}t^{4\alpha-1} + ...\right)
\end{align}
and the leading order is $(1/d)t^{4\alpha}\sum_{P\in\mathcal{P}_m}h_P^{2\alpha}=t^{4\alpha}\zeta_{\alpha}(h)$, further we have 
\begin{align}
    \zeta_{\alpha}(h)\ge \frac{1}{d}(h^{\dagger} h)^{2\alpha}=\frac{1}{d}||h||^{4\alpha}>0
\end{align}
so $\phi(t)$ cannot be linear function, and $\phi''(t)$ cannot always be $0$ in the interval $[0,1]$, therefore 
$\zeta_{\alpha}(f+th)$ must be strictly convex. More precisely, if we denote $\phi(t)=\zeta_{\alpha}(f+t(h-f))$, 
\begin{align}
    \zeta_{\alpha}(h) - \zeta_{\alpha}(f) - \nabla \zeta_{\alpha}(f)\cdot (h-f) &= \phi(1)-\phi(0)-\phi'(0) \\
    &=\int_0^1 \left[\phi'(t)-\phi'(0)\right] \mathrm{d} t\\
    &=\int_0^1\int_0^t \phi''(s)\mathrm{d} s\\
    &=\int_0^1(1-s)\phi''(s)\mathrm{d}s >0
\end{align}
so in conclusion we obtain 
\begin{align}
    \zeta_{\alpha}(h)> \zeta_{\alpha}(f)+\nabla\zeta_{\alpha}(f)\cdot(h-f), \quad h, f\in \mathbb{R}^d, h\neq f. \label{convex_property}
\end{align}

Next, we need to consider the off-diagonal elements will not contribute first and second order terms. 
For a general non-diagonal $H$, we merge two indices into one index, 
\begin{align}
    H_{a,b}\to H_x, \quad x=(a,b)\in \mathbb{F}_2^{2m},
\end{align}
and we denote $\phi(x)=H_{a,b}$, we further define 
\begin{align}
    B_s(k) = \sum_{x\in\mathbb{F}_2^{2m}}(-1)^{k\cdot x}\phi(x)\phi^*(x\oplus s)
\end{align}
then we have 
\begin{align}
    \zeta_{\alpha}(H)&=\frac{1}{d^2}\sum_{s,k\in\mathbb{F}_2^{2m}}|B_s(k)|^{2\alpha}\\
    &= \frac{1}{d^2}\sum_s \sum_{x_1\oplus...\oplus x_{2\alpha}=0}\left[
        \prod_{r=1}^{\alpha}\phi(x_r)\phi^*(x_r\oplus s)\right]
        \left[\prod_{r=\alpha+1}^{2\alpha}\phi^*(x_r)\phi(x_r\oplus s)\right].
\end{align}
We deonte $\mathcal{L}:=\{(a,a)|a\in \mathbb{F}_2^m\}$, then $x\in\mathcal{L}$ means $\phi(x)$ is a 
diagonal element of $H$. We denote $h=\operatorname{diag} H$, then we have the following 
conclusion, 
\begin{align}
    \zeta_{\alpha}(H)=\zeta_{\alpha}(h) + O(||H-h||^4). \label{nondiag_factor}
\end{align}
This is because, when $s\notin\mathcal{L}$, the two indices $x_r, x_r\oplus s$ cannot all be within $\mathcal{L}$, 
and they at least contribute $2\alpha\ge 4$ non-diagonal terms. When $s\in\mathcal{L}$, 
the two indices $x_r, x_r\oplus s$ either is both in $\mathcal{L}$, or both outside $\mathcal{L}$, 
so in this case, they also at least contribute $4$ non-diagonal terms. 
Therefore, for each non-diagonal elements, it at least contribute four terms, so we have 
Eq.~\eqref{nondiag_factor}. 

Finally we consider the general situation, for any hermitian matrix, we can write 
\begin{align}
    H(t) = e^{iKt}\Sigma e^{-iKt}=\Sigma + it[K,\Sigma] -\frac{t^2}{2}[K,[K,\Sigma]] + O(t^3),
\end{align}
bacause $\Sigma$ is diagonal, so $[K,\Sigma]_{ii}=0$, and we also have 
\begin{align}
    (K^2)_{ii} = \sum_j K_ij K_{ji}=\sum_j|K_{ij}|^2
\end{align}
therefore we have 
\begin{align}
    (K\Sigma K)_{ii}&=\sum_j f_j|K_{ij}|^2, \\
    (K^2\Sigma)_{ii} = (\Sigma K^2)_{ii} &= f_i\sum_j|K_{ij}|^2 
\end{align}
and put these in $H(t)$, we obtain 
\begin{align}
    H_{ii}(t) = f_i +  t^2\sum_j |K_{ij}|^2 (f_j-f_i) + O(t^3),\label{diag_term}
\end{align}
and for non-diagonal part, denote $E(t) := H(t) - h(t)=O(t)$ where $h(t)=\operatorname{diag}\ H(t)$, and from 
Eq.~\eqref{nondiag_factor}, we have 
\begin{align}
    \zeta_{\alpha}(H(t)) - \zeta_{\alpha}(h(t)) = O(||E(t)||^4)=O(t^4). 
\end{align}

Denote $g_i = \partial_i \zeta_{\alpha}(f)$ and $v_i = \sum_j|K_{ij}|^2(f_j-f_i)$, 
by using Taylor expansion and from Eq.~\eqref{diag_term} we know $h(t)=f+ t^2 v + O(t^3)$, we have 
\begin{align}
    \zeta_{\alpha}(h(t)) &= \zeta_{\alpha}(f) + \sum_i g_i [h_i(t)-f] + O(||h(t)-f||^2)\\
    &= \zeta_{\alpha}(f) + t^2\sum_i g_i v_i + O(t^3).
\end{align}
Put everything together, we have 
\begin{align}
    \zeta_{\alpha}(H(t)) = \zeta_{\alpha}(f) + t^2\sum_{i,j}g_i |K_{ij}|^2 (f_j-f_i) + O(t^3),
\end{align}
for $i<j$, the pair sum is 
\begin{align}
    g_i|K_{ij}|^2(f_j-f_i) + g_j|K_{ji}|^2(f_i-f_j)  = -(f_i-f_j)(g_i-g_j)|K_{ij}|^2,
\end{align}
if we define $c_{ij}^{(\alpha)} = (f_i-f_j)(g_i-g_j)$, we obtain 
\begin{align}
    \zeta_{\alpha}(e^{itK}\Sigma e^{-itK}) = \zeta_{\alpha}(\Sigma) - t^2\sum_{i<j}c_{ij}^{\alpha}|K_ij|^2 + O(t^3), 
\end{align}
which means
\begin{align}
    \frac{\mathrm{d}\zeta_{\alpha}(H(t))}{\mathrm{d} t}\Bigg|_{t=0} &= 0,\\
    \frac{\mathrm{d}^2\zeta_{\alpha}(H(t))}{\mathrm{d} t^2}\Bigg|_{t=0} &= 
    -2\sum_{i<j}c_{ij}^{(\alpha)}|K_{ij}|^2. \label{second_order}
\end{align}

If we assume that for $i<j$, $f_i\neq f_j$, then we denote $w=\pi_{ij} f$, then $w\neq f$, and 
\begin{align}
    w_k = \begin{cases}
        f_i, & k = j,\\
        f_j, & k = i,\\
        f_k, & \text{otherwise}. 
    \end{cases}
\end{align}
Now $w-f$ only have two non-zero terms, $(w-f)_i=f_j-f_i$ and $(w-f)_{j} = f_i-f_j$, 
so from Eq.~\eqref{convex_property}, we can write 
\begin{align}
    \zeta_{\alpha}(w)-\zeta_{\alpha}(f)>\nabla\zeta_{\alpha}(f)\cdot (w-f) = -c_{ij}^{(\alpha)}, 
\end{align}
which implies 
\begin{align}
    c_{ij}^{(\alpha)}>\zeta_{\alpha}(f)-\zeta_{\alpha}(\pi_{ij} f).
\end{align}
If $f$ is already in descending order, i.e., $f_i>f_j$ for any $i<j$, then we always have 
$\zeta_{\alpha}(f)-\zeta_{\alpha}(\pi_{ij} f)\ge0$, which means, 
\begin{align}
    c_{ij}^{(\alpha)}>\zeta_{\alpha}(f)-\zeta_{\alpha}(\pi_{ij} f)\ge 0.
\end{align}
Therefore, Eq.~\eqref{second_order} is always non-positive. 
So reference state is always a local optimal point for nonlocal SRE optimization problem.

\section{Bounds on nonlocal SRE}
\subsection{Upper bound}
We now prove that 
\begin{align}
    M_{\alpha}^{\mathrm{Sch}}\le 2 S_{\alpha},
\end{align}
where $S_{\alpha}=(1-\alpha)^{-1}\log_2 P_{\alpha}$ is the R\'enyi entanglement entropy, and we denote $P_{\alpha}=\sum_i\lambda_i^{\alpha}$. 
This upper bound for $\alpha=2$ was previously established in Refs.~\cite{Cao2025,Fabio2026,Gianpaolo2026}, and here we extend it to general $\alpha$. 

For $\alpha>1$, we have (using Jensen's inequality)
\begin{align}
    \frac{1}{d}\sum_{k}|A_s(k)|^{2\alpha} 
    \ge \left(\frac1d\sum_{k}\left|A_s(k)\right|^2\right)^{\alpha} \label{eq282}
\end{align}
and by Parseval identity one has 
\begin{align}
    \frac1d \sum_k |A_s(k)|^2=\sum_x f_s(x)^2 = \sum_x \lambda_x\lambda_{x\oplus s}
\end{align}
where we denote $f_s(x)\equiv \sqrt{\lambda_x\lambda_{x\oplus s}}$.
This is becasue 
\begin{align}
    \sum_k|A_s(k)|^2=\sum_k A_s(k)^2 &= \sum_k\left[\sum_x(-1)^{k\cdot x}f_s(x)\right]
    \left[\sum_y(-1)^{k\cdot y}f_s(y)\right]\\
    &= \sum_{k,x,y}(-1)^{k\cdot x}(-1)^{k\cdot y}f_s(x)f_s(y)\\
    \text{using } (-1)^{k\cdot x}(-1)^{k\cdot y}=(-1)^{k\cdot(x\oplus y)}\quad & = 
    \sum_{x,y}f_s(x)f_s(y)\sum_k(-1)^{k\cdot(x\oplus y)}\\
    \text{using } \sum_{k\in \mathbb{F}_2^m}(-1)^{k\cdot z}=d\delta_{z,0}\quad &=\sum_{x,y}f_s(x)f_s(y)d\delta_{x,y}\\
    &= d\sum_x f_s(x)^2
\end{align}
where the two equality can be simply proved,
\begin{align}
    (-1)^{k\cdot x}(-1)^{k\cdot y} &= \prod_i(-1)^{k_i x_i}\prod_i(-1)^{k_i y_i}=\prod_i(-1)^{k_i (x_i+y_i)}
    = \prod_i(-1)^{k_i (x_i\oplus y_i)}=(-1)^{k\cdot(x\oplus y)},\\
    \sum_{k\in \mathbb{F}_2^m}(-1)^{k\cdot z}&=\sum_{k\in \mathbb{F}_2^m}\prod_i (-1)^{k_i z_i}
    =\prod_{i=1}^m\left[\sum_{k_i=0}^1(-1)^{k_i z_i}\right]=\prod_{i=1}^m[1+(-1)^{z_i}]=2^m\delta_{z,0}=d\delta_{z,0}. 
\end{align}

So from Eq.~\eqref{eq282}, we have (when $\alpha>1$)
\begin{align}
    \frac{1}{d}\sum_{k}|A_s(k)|^{2\alpha} 
    \ge \left(\frac1d\sum_{k}\left|A_s(k)\right|^2\right)^{\alpha}
    =\left(\sum_x f_s(x)^2\right)^{\alpha}\ge \sum_x f_s(x)^{2\alpha}=\sum_x \lambda_x^{\alpha}\lambda_{x\oplus s}^{\alpha}
\end{align}
then 
\begin{align}
    \zeta_{\alpha}^{\mathrm{Sch}} = 
    \frac{1}{d}\sum_{s,k=0}^{2^m-1} |A_s(k)|^{2\alpha}\ge \sum_s\sum_x\lambda_x^{\alpha}\lambda_{x\oplus s}^{\alpha}
    =\left(\sum_x\lambda_x^{\alpha}\right)^2,
\end{align}
therefore 
\begin{align}
    \zeta_{\alpha}^{\mathrm{Sch}} \ge P_{\alpha}^2 &\Rightarrow \log_2\zeta_{\alpha}^{\mathrm{Sch}}\ge \log_2 P_{\alpha}^2\\
    &\Rightarrow M_{\alpha}^{\mathrm{Sch}}=\frac{1}{1-\alpha}\log_2\zeta_{\alpha}^{\mathrm{Sch}}\le \frac{2}{1-\alpha}\log_2 P_{\alpha}
    =2S_{\alpha}. 
\end{align}
So we prove that $M_{\alpha}^{\mathrm{Sch}}\le 2S_{\alpha}$ when $\alpha >1$. 

When $\alpha < 1$, we have 
\begin{align}
    \frac{1}{d}\sum_{k}|A_s(k)|^{2\alpha} 
    \le \left(\frac1d\sum_{k}\left|A_s(k)\right|^2\right)^{\alpha}
\end{align}
so in this case 
\begin{align}
    \zeta_{\alpha}^{\mathrm{Sch}} \le P_{\alpha}^2
\end{align}
but $1-\alpha$ is positive now, therefore one still have 
\begin{align}
     M_{\alpha}^{\mathrm{Sch}}=\frac{1}{1-\alpha}\log_2\zeta_{\alpha}^{\mathrm{Sch}}\le \frac{2}{1-\alpha}\log_2 P_{\alpha}
    =2S_{\alpha}. 
\end{align}

\subsection{Lower bound}
We now define generalized anti-flatness as 
\begin{align}
    \mathcal{A}_n = P_{2n-1}-P_n^2 = \sum_i\lambda_i^{2n-1}-\left(\sum_i\lambda_i^n\right)^2=\sum_{i<j}\lambda_i\lambda_j(\lambda_i^{n-1}-\lambda_j^{n-1})^2
\end{align}
where $P_n=\sum_i\lambda_i^n$. 
This is because 
\begin{align}
    \sum_{i<j}\lambda_i\lambda_j(\lambda_i^{n-1}-\lambda_j^{n-1})^2 &= \frac12\sum_{i,j}\lambda_i\lambda_j(\lambda_i^{n-1}-\lambda_j^{n-1})^2\\
    &= \frac12\sum_{i,j}\lambda_i\lambda_j\left(\lambda_i^{2n-2}+\lambda_j^{2n-2}-2\lambda_i^{n-1}\lambda_j^{n-1}\right)\\
    &= \frac{1}{2}\sum_{i,j}\lambda_i^{2n-1}\lambda_j + \lambda_i\lambda_j^{2n-1}-2\lambda_i^n\lambda_j^n\\
    &=\frac12\left(P_{2n-1}P_1+P_1P_{2n-1}-2P_n^2\right)\\
    &= P_{2n-1}-P_n^2. 
\end{align}
We have the following lower bound for nonlocal $\alpha$-SRE, 
\begin{align}
    M_{\alpha}^{\mathrm{Sch}}\ge \frac{1}{\alpha-1}\log_2\frac{1}{1-2\alpha\mathcal{A}_{\alpha}}. 
\end{align}
Now we try to prove it. 

Direct expanding $\zeta_{\alpha}^{\mathrm{Sch}}$, we have 
\begin{align}
    \zeta_{\alpha}^{\mathrm{Sch}}&=
    \frac{1}{d}\sum_{s}\sum_{k}\left[\sum_x(-1)^{k\cdot x}f_s(x)\right]^{2\alpha}\\
    &= \frac1d\sum_s\sum_k\prod_{i=1}^{2\alpha}\left[\sum_{x_i}(-1)^{k\cdot x_i}f_s(x_i)\right]\\
    &= \frac1d\sum_{s}\sum_{k}\sum_{x_1,...,x_{2\alpha}}(-1)^{k\cdot (x_1\oplus...\oplus x_{2\alpha})}
    f_s(x_1)f_s(x_2)...f_s(x_{2\alpha})\\
    &= \sum_s\sum_{x_1,...,x_{2\alpha}}\left[\frac1d\sum_k (-1)^{k\cdot(x_1\oplus...\oplus x_{2\alpha})}\right]
    f_s(x_1)f_s(x_2)...f_s(x_{2\alpha})\\
    &= \sum_s\sum_{\substack{x_1,x_2,...,x_{2\alpha}\\x_1\oplus...\oplus x_{2\alpha}=0}}
    \prod_{i=1}^{2\alpha}f_s(x_i)\label{eq307}
\end{align}
and now we define new $2\alpha$ variables
\begin{align}
    y_i = \begin{cases}
        x_i, & i\in [1,2\alpha-1]\\
        x_{i}\oplus s, & i=2\alpha
    \end{cases}
\end{align}
so the summation is 
\begin{align}
    S = y_1\oplus y_2\oplus...\oplus y_{2\alpha}= s.
\end{align}
We also define 
\begin{align}
    W(\boldsymbol{y})=\prod_{i=1}^{2\alpha}\lambda_{y_i}=\lambda_{y_1}\lambda_{y_2}...\lambda_{y_{2\alpha}},
\end{align}
and the $T$ transformation
\begin{align}
    T\boldsymbol{y}=(y_1\oplus S,...,y_{2\alpha}\oplus S)
\end{align}
so the $\zeta_{\alpha}^{\mathrm{Sch}}$ in Eq.~\eqref{eq307} can be written as 
\begin{align}
    \zeta_{\alpha}^{\mathrm{Sch}} = \sum_{\boldsymbol{y}}\sqrt{
        W(\boldsymbol{y})W(T\boldsymbol{y})
    }
\end{align}
note that $T^2=\operatorname{id}$, so
\begin{align}
    \sum_{\boldsymbol{y}}W(\boldsymbol{y})=\sum_{\boldsymbol{y}}W(T\boldsymbol{y})=\left(\sum_i\lambda_i\right)^{2\alpha}=1
\end{align}
therefore we have 
\begin{align}
    1-\zeta_{\alpha}^{\mathrm{Sch}} &= \frac12\left[\sum_{\boldsymbol{y}}\sqrt{W(\boldsymbol{y})}\sqrt{W(\boldsymbol{y})}
    +\sum_{\boldsymbol{y}}\sqrt{W(T\boldsymbol{y})}\sqrt{W(T\boldsymbol{y})}
    -2\sum_{\boldsymbol{y}}\sqrt{W(\boldsymbol{y})}\sqrt{W(T\boldsymbol{y})}\right]\\
    &= \frac12\sum_{\boldsymbol{y}}\left(\sqrt{W(\boldsymbol{y})}-\sqrt{W(T\boldsymbol{y})}\right)^2\ge 0.
\end{align}
So for every $\boldsymbol{y}$ above, the difference is non-negative. 
We can only consider one special situation. 

We now only consider $\boldsymbol{y}=(a,a,...,a,b)$ with $a\neq b$. 
So now 
\begin{align}
    S = a\oplus a...\oplus a\oplus b = \left(\oplus_{i=1}^{2\alpha-1}a\right) \oplus b=a\oplus b,
\end{align}
therefore $T$ is 
\begin{align}
    Ta = b,\quad Tb=a. 
\end{align}
So 
\begin{align}
    \left(\sqrt{W(\boldsymbol{y})}-\sqrt{W(T\boldsymbol{y})}\right)^2&=\lambda_a^{2\alpha-1}\lambda_b+\lambda_b^{2\alpha-1}\lambda_a-2\lambda_a^{\alpha}\lambda_b^{\alpha}\\
    &= \lambda_a\lambda_b(\lambda_a^{\alpha-1}-\lambda_b^{\alpha-1})^2.
\end{align}
For unorder pair $(a,b)$, there are $2\alpha$ choice to put $b$ and consider the exchange of $a$ and $b$, so there are 
$4\alpha$ contribution of partition $[2\alpha-1,1]$, denoted $\Delta_{2\alpha-1|1}$, 
\begin{align}
    \Delta_{2\alpha-1|1} = 4\alpha\sum_{a<b}\frac12\lambda_a\lambda_b(\lambda_a^{\alpha-1}-\lambda_b^{\alpha-1})^2
\end{align}
and 
\begin{align}
   \mathcal{A}_{\alpha}= P_{2\alpha-1}-P_{\alpha}^2 = \sum_{a<b}\lambda_a\lambda_b(\lambda_a^{\alpha-1}-\lambda_b^{\alpha-1})^2
\end{align}
so we obtain
\begin{align}
    \Delta_{2\alpha-1|1} =2\alpha \mathcal{A}_{\alpha}
\end{align}
and consequently
\begin{align}
    1-\zeta_{\alpha}^{\mathrm{Sch}}\ge \Delta_{2\alpha-1|1}=2\alpha \mathcal{A}_{\alpha}
\end{align}
so we have 
\begin{align}
    M_{\alpha}^{\mathrm{Sch}}\ge \frac{1}{\alpha-1}\log_2\frac{1}{1-2\alpha\mathcal{A}_{\alpha}}.
\end{align}

\section{Nonlocal SRE of exponentially and algebraically decaying spectra}
\subsection{Exponentially decaying spectrum}
Suppose the Schmidt rank is $N=2^m$, and consider the following spectrum $\{\lambda_x\}$, 
\begin{align}
    \lambda_x = C_N q^x,\quad x = 0,1,...,N-1, \quad 0<q<1, 
\end{align}
where $C_N$ is the normalization factor, 
\begin{align}
    C_N\sum_{x=0}^{N-1}\lambda_x = 1\Rightarrow C_N = \frac{1-q}{1-q^N},
\end{align}
and therefore, 
\begin{align}
    \lambda_x = \frac{1-q}{1-q^N} q^x. 
\end{align}

We now calculate $P_{\alpha}\equiv \sum_x\lambda_x^{\alpha}$. For exponentially decaying spectrum, we have 
\begin{align}
    P_{\alpha} = (C_{N})^{\alpha} \sum_{x=0}^{N-1} q^{\alpha x} = 
    \left(\frac{1-q}{1-q^N}\right)^{\alpha}\frac{1-q^{\alpha N}}{1-q^{\alpha}}.
\end{align}
From this, the entropy and anti-flatness can easily obtained, 
\begin{align}
    S_{\alpha} &= \frac{1}{1-\alpha}\log_2 P_{\alpha} = 
    \frac{1}{\alpha-1}\left(\alpha\log_2\frac{1-q^N}{1-q} - \log_2\frac{1-q^{\alpha N}}{1-q^{\alpha}}\right),\\
    \mathcal{A}_2 &= P_3-P_2^2 = \frac{(1-q)^2(q-q^N)(1-q^{N+1})}{(1+q)^2(1+q+q^2)(1-q^N)^2},
\end{align}
and we can get a bound estimate of nonlocal SRE now, due to $2S_2\ge M_2^{\mathrm{Sch}}\ge-\log_2(1-4\mathcal{A}_2)$, 
in this situation we get
\begin{align}
    -\log_2\left[1-\frac{4(1-q)^2(q-q^N)(1-q^{N+1})}{(1+q)^2(1+q+q^2)(1-q^N)^2}\right] \le 
    M_2^{\mathrm{Sch}} \le 
    4\log_2\frac{1-q^N}{1-q}-2\log_2\frac{1-q^{2N}}{1-q^2}. 
\end{align}

For exponentially decaying spectrum, we can analytically obtain its $M_2^{\mathrm{Sch}}$. 
If we use the binary representation of the label $x$, 
\begin{align}
    x = \sum_{j=0}^{m-1}2^j x_j, \quad x_j \in \{0, 1\}, 
\end{align}
we have 
\begin{align}
    \lambda_x = \frac{1-q}{1-q^{2^m}}q^x = \frac{1-q}{1-q^{2^m}}q^{\sum_{j=0}^{m-1}2^j x_j}
     = \frac{1-q}{1-q^{2^m}}\prod_{j=0}^{m-1} q^{2^j x_j}, 
\end{align}
note that 
\begin{align}
    (1-q^{2^m}) &= (1-q^{2^{m-1}})(1+q^{2^{m-1}})\\
    &= (1-q^{2^{m-2}})(1+q^{2^{m-2}})(1+q^{2^{m-1}})\\
    & ...\\
    &= (1-q)\prod_{j=0}^{m-1}(1+q^{2^j}),
\end{align}
therefore, if we denote $r_j = q^{2^j}$, 
\begin{align}
    \lambda_x = \prod_{j=0}^{m-1}\frac{q^{2^j x_j}}{1+q^{2^j}} = \prod_{j=0}^{m-1}\frac{r_j^{x_j}}{1+r_j},
\end{align}
which tells us that the $N=2^m$ spectrum can be factorized as the tensor product of $m$ binary tuple:
\begin{align}
    \left(\frac{1}{1+r_j}, \frac{r_j}{1+r_j}\right). 
\end{align}
For one tuple we have, 
\begin{align}
    m_2^{\mathrm{Sch}} \left(\frac{1}{1+r}, \frac{r}{1+r}\right) = m_2^{\mathrm{Sch}}(r) = 
    \log_2\frac{(1+r)^4}{1+14r^2+r^4}, 
\end{align}
then for the whole spectrum, we have 
\begin{align}
    M_2^{\mathrm{Sch}} = \sum_{j=0}^{m-1}m_2^{\mathrm{Sch}}(r_j) 
    = \sum_{j=0}^{m-1}\log_2\frac{(1+q^{2^j})^4}{1+14q^{2^{j+1}}+q^{2^{j+2}}}.
\end{align}

We now discuss the results. 
In the $N\to+\infty$ limit we have 
\begin{align}
    S_{\alpha}(q,N=+\infty) &= \frac{\alpha}{\alpha-1}\log_2\frac{1-q^{\alpha}}{1-q}\\
    \mathcal{A}_2(q, N=+\infty) &= \frac{q(1-q)^2}{(1+q)^2(1+q+q^2)}.
\end{align}
The entropy and anti-flatness remain finite in the $N\to+\infty$ limit when $q\in(0, 1)$. 
The behaviour of nonlocal SRE is more interesting, we have 
\begin{align}
    M_2^{\mathrm{Sch}}(q,2N)-M_2^{\mathrm{Sch}}(q,N)
     = m_2^{\mathrm{Sch}}(q^N),
\end{align}
because 
\begin{align}
    m_2^{\mathrm{Sch}}(x) = \log_2\frac{(1+x)^4}{1+14x^2+x^4}
    \sim \frac{4x}{\ln 2}-\frac{16 x^2}{\ln 2} + \frac{4 x^3}{3\ln 2}+O(x^4),\quad x \to 0,
\end{align}
so 
\begin{align}
    M_2^{\mathrm{Sch}}(q, 2N)-M_2^{\mathrm{Sch}}(q, N) \sim \frac{4}{\ln 2}q^N.
\end{align}
we can also get 
\begin{align}
    M_2^{\mathrm{Sch}}(q, +\infty) - M_2^{\mathrm{Sch}}(q, N) 
    &= \left[M_2^{\mathrm{Sch}}(q, 2N)- M_2^{\mathrm{Sch}}(q, N)\right]\\
    &+ \left[M_2^{\mathrm{Sch}}(q, 4N)- M_2^{\mathrm{Sch}}(q, 2N)\right]
    + \left[M_2^{\mathrm{Sch}}(q, 8N)- M_2^{\mathrm{Sch}}(q, 4N)\right]
    + ...\\
    &= \sum_{\ell=0}^{+\infty}m_2^{\mathrm{Sch}}(q^{N 2^{\ell}})
    \sim \frac{4}{\ln 2}\sum_{\ell}^{+\infty}  q^{N2^{\ell}}\sim \frac{4}{\ln 2}q^N, 
\end{align}
which is 
\begin{align}
    M_2^{\mathrm{Sch}}(q, +\infty) - M_2^{\mathrm{Sch}}(q, N) =\frac{4}{\ln 2} q^N + O(q^{2N}). 
\end{align}

We also discuss the limit of $q\to 0$ and $q\to 1$. 
In the weakly entangled regime $q\to 0$, 
we have 
\begin{align}
     M_2^{\mathrm{Sch}}(q\to 0, N)&=m_2^{\mathrm{Sch}}(q)+m_2^{\mathrm{Sch}}(q^2)+...\\
     &\sim \frac{4q}{\ln 2}+O(q^2). 
\end{align}
And at this time we also have 
\begin{align}
    \mathcal{A}_2(q\to 0, N)\sim q+O(q^2), \\
    S_{2}(q\to 0, N)\sim \frac{2q}{\ln 2}+O(q^3). 
\end{align}
Therefore in this regime $q\to 0$, we have 
\begin{align}
    M_2^{\mathrm{Sch}}\simeq\frac{4}{\ln 2}\mathcal{A}_2\simeq 2S_2,\quad q\to 0, N\ge 2. 
\end{align}

For $q\to 1$ limit, the situation is much more interesting. We denote $q=1-\epsilon$, and in this limit $\epsilon\to 0$, 
the spectrum tends to be flat, so $S_2\sim \log_2 N$, $M_2^{\mathrm{Sch}}\to 0$ and $\mathcal{A}_2\to 0$. 
We now calculate the leading behaviour, for anti-flatness
\begin{align}
    \mathcal{A}_2(q\to 1,N) \sim \frac{(N^2-1)}{12N^2}\epsilon + O(\epsilon^3), 
\end{align}
for $S_2$, 
\begin{align}
    S_2(q\to 1, N)\sim \log_2 N - \frac{(N^2-1)}{12\ln 2}\epsilon^2 + O(\epsilon^3),
\end{align}
for nonlocal SRE
\begin{align}
    M_2^{\mathrm{Sch}}(q\to 1, N) &\sim \sum_{j=0}^{m-1}\left(\frac{4^j \epsilon^2}{4\ln 2} + O(\epsilon^3)\right)
    \sim \left(\sum_{j=0}^{m-1}4^j\right)\frac{\epsilon^2}{4\ln 2}\\
    &\sim \frac{4^m-1}{3}\frac{\epsilon^2}{4\ln 2}=\frac{N^2-1}{12\ln 2}\epsilon^2.
\end{align}
In conclusion, in this limit $q\to 1$, we have the relations
\begin{align}
    M_2^{\mathrm{Sch}}\simeq\frac{N^2}{\ln 2}\mathcal{A}_2\simeq\log_2N - S_2. 
\end{align}
It is very interesting to note that when $N$ is large and the spectrum is nearly flat, 
$\mathcal{A}_2$ is no longer suitable to describe nonlocal SRE, becuase there is a $N^2$ factor. 

Finally we note that 
\begin{align}
    \lim_{N\to+\infty}\lim_{q\to 1} M_2^{\mathrm{Sch}}(q,N)=
    0\neq \lim_{q\to 1}\lim_{N\to+\infty} M_2^{\mathrm{Sch}}(q,N),
\end{align}
because we assume $N=2^m$ above. 

\subsection{Algebraically decaying spectrum}
We now study the algebraically decaying spectrum $\{\lambda_r\}$,
\begin{align}
    \lambda_r = \frac{r^{-p}}{H(N,p)},\quad r = 1,2,...,N,\quad p>0
\end{align}
where $H(N,p)$ (or equivalently $H_{N,p}$) is the normalization factor
\begin{align}
    H(N,p) = \sum_{r=1}^N\frac{1}{r^p}. 
\end{align}

We also define the $\alpha$ order moments
\begin{align}
    P_{\alpha}(N,p)=\sum_{r=1}^{N}\lambda_r^{\alpha} = \frac{1}{H^{\alpha}(N,p)}\sum_{r=1}^{N} r^{-\alpha p}
    = \frac{H(N,\alpha p)}{H^{\alpha}(N, p)}. 
\end{align}
In the following, we may use $H(N,p)$ or $H_{N,p}$. They are interchangable. 
So for R\'enyi entropies we have 
\begin{align}
    S_{\alpha}(N,p) = \frac{1}{1-\alpha}\log_2\frac{H(N,\alpha p)}{H^{\alpha}(N,p)},
\end{align}
and specifically when $\alpha=2$ we obtain 
\begin{align}
    S_2(N,p) = 2\log_2 H(N,p) - \log_2 H(N,2 p). 
\end{align}

We can also obtain exact expression for anti-flatness $\mathcal{A}_2(N,p)$, 
\begin{align}
    \mathcal{A}_2(N,p) = P_3-P_2^2 = \frac{H(N,p)H(N,3p)-H^2(N,2p)}{H^4(N,p)}
\end{align}
therefore we have at least a sound estimate of nonlocal SRE, 
\begin{align}
    -\log_2\left(1-4\frac{H(N,p)H(N,3p)-H^2(N,2p)}{H^4(N,p)}\right)\le M_2^{\mathrm{Sch}}(N,p)\le 2
    \left[2\log_2 H(N,p) - \log_2 H(N,2 p)\right]. 
\end{align}

Now we want to find a more finer bound. Suppose $N=2^m$, 
and label the spectrum
\begin{align}
    \lambda_x = \frac{1}{H(N,p)}\frac{1}{(1+x)^p},\quad x = 0,1,..., N-1. 
\end{align}
We now define the dyadic shell, 
\begin{align}
    \mathcal{S}_j = \{2^j, ..., 2^{j+1}-1\}, \quad j = 0,..., m-1,
\end{align}
where each shell has $|\mathcal{S}_j|=2^j$ elements. 
If $a, b, c, d \in \mathcal{S}_j$ then their $j$-th binary bit is $1$ so the $j$-th bit of $s=a\oplus b\oplus c\oplus d$ 
must be $1\oplus 1\oplus 1\oplus 1=0$. So $a\oplus s,b\oplus s,c\oplus s,d\oplus s\in S_j$. 
For element $x\in S_j$, we have $x+1\le 2^{j+1}$, 
therefore 
\begin{align}
    \lambda_x\ge \frac{1}{H(N,p)}\frac{1}{2^{p(j+1)}}, 
\end{align}
and consequently 
\begin{align}
    \sqrt{\lambda_a\lambda_b\lambda_c\lambda_d
    \lambda_{b\oplus c\oplus d}\lambda_{a\oplus c\oplus d}\lambda_{a\oplus b\oplus d}\lambda_{a\oplus b\oplus c}}
    \ge\left[\frac{2^{-p(j+1)}}{H(N,p)}\right]^4, 
\end{align}
and because there are $(2^j)^4$ terms in the summation, 
\begin{align}
    \zeta_2^{\mathrm{Sch}}=2^{-M_2^{\mathrm{Sch}}}\ge\frac{2^{-4p}}{H^4(N,p)}\sum_{j=0}^{m-1}2^{4j(1-p)}, 
\end{align}
finally we have an another bound for nonlocal SRE
\begin{align}
    M_2^{\mathrm{Sch}}(N,p)\le -\log_2\left[
        \frac{2^{-4p}}{H^4(N,p)}\sum_{j=0}^{m-1}2^{4j(1-p)}
    \right] = 4p+4\log_2 H(N,p)-\log_2 G_m(p) \label{eq93}, 
\end{align}
where $G_m(p)$ is defined as 
\begin{align}
    G_m(p) = \sum_{j=0}^{m-1}2^{4j(1-p)} = \begin{cases}
        m=\log_2N & p=1,\\
        \displaystyle \frac{N^{4(1-p)}-1}{2^{4(1-p)}-1} & p\neq 1.
    \end{cases}
\end{align}
In summary we have 
\begin{align}
    M_2^{\mathrm{Sch}}(N,p)\le \min\Big\{
        2\left[2\log_2 H(N,p) - \log_2 H(N,2 p)\right], 
        4p+4\log_2 H(N,p)-\log_2 G_m(p)
    \Big\}.
\end{align}

Now we discuss the limit $N\to +\infty$. 
We first introduce the scaling behaviour of $H(N,s)=\sum_{r=1}^N r^{-s}$, 
\begin{align}
    H(N\to+\infty, s)=\begin{cases}
        \displaystyle \zeta(s) - \frac{1}{s-1}N^{-(s-1)} + O(N^{-s}), & s>1,\\
        \displaystyle \ln N+ \gamma + O(N^{-1}), & s=1,\\
        \displaystyle \frac{1}{1-s}N^{1-s} + \zeta(s) + O(N^{-s}), & 0<s<1.
    \end{cases}
\end{align}
Now we deal with the case $p>1$. 
For $S_2$ we have 
\begin{align}
    S_2(N\to+\infty, p) = 2\log_2\zeta(p)-\log_2\zeta(2p), 
\end{align}
which is a finite value, 
and for anti-flatness we have 
\begin{align}
    \mathcal{A}_2(N\to+\infty, p) = \frac{\zeta(p)\zeta(3p)-\zeta(2p)^2}{\zeta(p)^4},
\end{align}
which is also a finite value. 
So the nonlocal SRE must be a finite value because we have 
\begin{align}
    -\log_2\left[1-4\frac{\zeta(p)\zeta(3p)-\zeta(2p)^2}{\zeta(p)^4}\right]\le M_2^{\mathrm{Sch}}(N\to+\infty,p)\le 2[2\log_2\zeta(p)-\log_2\zeta(2p)], 
\end{align}
therefore at this situation we have $M_2^{\mathrm{Sch}}(N\to+\infty,p)=\Theta(1)$. 

When $p=1$, we first have 
\begin{align}
    S_2(N\to+\infty,p=1) &= 2\log_2\left[\ln N+\gamma + O(N^{-1})\right]-\log_2\zeta(2)\sim 2\log_2\ln N+O(1), \\
    \mathcal{A}_2(N\to+\infty,p=1) &= \frac{\left[\ln N+\gamma + O(N^{-1})\right]\zeta(3)-\zeta(2)^2}{\left[\ln N+\gamma + O(N^{-1})\right]^4}
    \sim \frac{\zeta(3)}{(\ln N)^3}, 
\end{align}
we note that in this case $S_2\to \infty$ while $\mathcal{A}_2\to 0$, specifically for $\mathcal{A}_2$
\begin{align}
    M_2^{\mathrm{Sch}}(N\to+\infty,p=1)\ge \frac{4\mathcal{A}_2}{\ln 2} = \frac{4\zeta(3)}{\ln 2(\ln N)^3}. 
\end{align}
At this stage, we can only get 
\begin{align}
    M_2^{\mathrm{Sch}}(N\to+\infty,p=1) \le 4\log_2\ln N.
\end{align}
From Eq.~\eqref{eq93}, we can obtain a slightly stricter bound, 
\begin{align}
    M_2^{\mathrm{Sch}}(N\to+\infty,p=1) \le 4 + 4\log_2\left[\ln N+\gamma+O(N^{-1})\right]-\log_2 \log_2 N
    = 3\log_2\ln N + O(1). 
\end{align}
Therefore, at least we have 
\begin{align}
    M_2^{\mathrm{Sch}}(N\to+\infty,p=1) \le 3\log_2\ln N. 
\end{align}

Finally we discuss the most complicated case, $p\in (0, 1)$. 
For $S_2$ we have 
\begin{align}
    S_2(N\to+\infty,p)=\begin{cases}
        \displaystyle 2(1-p)\log_2 N , & p\in(1/2,1)\\
        \displaystyle \log_2 N - \log_2 \ln N + 2+ O(1), & p=1/2\\
        \displaystyle \log_2 N - \log_2 \frac{(1-p)^2}{1-2p}+O(1), & p\in(0, 1/2). 
    \end{cases}
\end{align}
So in this case, all $S_2$ scales as 
\begin{align}
    S_2(N\to+\infty,p)\sim \log_2 N. 
\end{align}

For anti-flatness we have 
\begin{align}
    \mathcal{A}_2(N\to+\infty,p)\simeq\begin{cases}
        \displaystyle \zeta(3p)(1-p)^3 N^{-3(1-p)}, & p\in (1/3,1),\\
        \displaystyle \frac{8}{27}\frac{\ln N}{N^2}, & p=1/3,\\
        \displaystyle \frac{p^2(1-p)^3}{(1-2p)^2(1-3p)} N^{-2}, & p\in (0, 1/3). 
    \end{cases}
\end{align}
Similarly, in this case $p\in(0,1)$
\begin{align}
    \mathcal{A}_2(N\to+\infty,p)\to 0. 
\end{align}

For nonlocal SRE we obtain
\begin{align}
    M_2^{\mathrm{Sch}}(N\to+\infty,p)&\le 4p+4\log_2 H(N,p)-\log_2 G_m(p)\\
    &= 4p+4\log_2 \left[
        \frac{1}{1-p}N^{1-p} + \zeta(p) + O(N^{-p})
    \right]-\log_2 \frac{N^{4(1-p)}-1}{2^{4(1-p)}-1}\\
    &\simeq 4p + \log\left[\frac{N^{4(1-p)}}{(1-p)^4}\frac{2^{4(1-p)}-1}{N^{4(1-p)}-1}\right]\\
    &\simeq \log_2\frac{2^{4}-2^{4p}}{(1-p)^4}\equiv\log_2 c_p. 
\end{align}
So we obtain 
\begin{align}
    M_2^{\mathrm{Sch}}(N\to+\infty,p) \le \log_2 c_p + O(1),
\end{align}
so when $q\in (0, 1)$, 
\begin{align}
    M_2^{\mathrm{Sch}}(N\to+\infty,p) = O(1). 
\end{align}

Now we summaize the results in the following. This is the Table. I in the main text.  
\begin{table}[H]
    \centering
    \begin{tabular}{cccc}
    \hline
    $p$            & $S_2$                      & $\mathcal{A}_2$        & $M_2^{\mathrm{Sch}}$ \\ 
    \hline
    $(1, +\infty)$ & $O(1)$                     & $O(1)$               & $O(1)$                           \\
    $1$            & $2\log_2\ln N + O(1)$      & $\zeta(3)/(\ln N)^3$ & $O(\log_2\ln N)$                 \\
    $(1/2, 1)$     & $2(1-p)\log_2 N + O(1)$    & $N^{-3(1-p)}$        & $O(1)$                           \\
    $1/2$          & $\log_2N-\log_2\ln N+O(1)$ & $N^{-3/2}$           & $O(1)$                           \\
    $(1/3, 1/2)$   & $\log_2 N+O(1)$            & $N^{-3(1-p)}$        & $O(1)$                           \\
    $1/3$          & $\log_2 N+O(1)$            & $N^{-2}\ln N$        & $O(1)$                           \\
    $(0, 1/3)$     & $\log_2 N+O(1)$            & $N^{-2}$             & $O(1)$                           \\
    \hline
    \end{tabular}
\end{table}

\section{The sandwich construction}
\subsection{Bounds on shell flat spectrum}
We define the following sets $B_j$ and $S_j$, 
\begin{align}
    B_j &= \{0,..., 2^j-1\} \simeq \mathbb{F}_2^j, \\
    S_j &= \{2^j,..., 2^{j+1}-1\},\\
    B_{j+1} &= B_j \cup S_j,
\end{align}
and we consider the shell flat spectrum, i.e., 
\begin{align}
    \lambda_x=\frac{\mu_j}{2^j},\quad x\in S_j,
\end{align}
where $\mu_j=\sum_{x\in S_j}\lambda_x$ is the total weight of $S_j$. 
We define the function $g_j(x)=\sqrt{\lambda_x}$ for $\forall x \in B_j$, and we also define 
\begin{align}
    g_{j+1}(x) = \begin{cases}
        g_j(x), & x\in B_j, \\
        \sqrt{\mu_j/2^j}, & x\in S_j. 
    \end{cases}
\end{align}

We define two quantities $\Gamma_j(s)$ and $H_j$ as 
\begin{align}
    \Gamma_j(s) = \sum_{x\in B_j}g_j(x) g_j(x\oplus s), 
\end{align}
and 
\begin{align}
    H_j=\frac{1}{2^j}\left(\sum_{x\in B_j}g_j(x)\right)^2, 
\end{align}
and we further define for integer $p$ that 
\begin{align}
    R_j^{(p)}=\frac{1}{2^j}\sum_{s\in B_j}\left[\Gamma_j(s)\right]^p.
\end{align}
Note that we have $R_j^{(1)}=H_j$. This is because 
\begin{align}
    R_j^{(1)} = \frac{1}{2^j}\sum_{s\in B_j}\Gamma_j(s)=
    \frac{1}{2^j}\sum_{s\in B_j}\sum_{x\in B_j}g_j(x) g_j(x\oplus s)
    =\frac{1}{2^j}\left(\sum_{x\in B_j}g_j(x)\right)^2=H_j. 
\end{align}
Now if we add a new flat spectrum $s\to \sigma s \in \{0s,1s\}$, when $\sigma=0$, 
\begin{align}
    \Gamma_{j+1}(0s)=\Gamma_j(s)+\mu_j
\end{align} 
when $\sigma=1$, 
\begin{align}
    \Gamma_{j+1}(1s)=2q_j\sum_{x\in B_j} g_j(x)=2\sqrt{\mu_j H_j},
\end{align}
in conclusion we now have 
\begin{align}
    \Gamma_{j+1}=\Gamma_j+\mu_j+2\sqrt{\mu_j H_j}
\end{align}
so for $R_j^{(p)}$ we have 
\begin{align}
    R_{j+1}^{(p)} &= \frac{1}{2^{j+1}}\sum_{s\in B_j}
        \Gamma_{j+1}(0s)^p+\Gamma_j(1s)^p\\
    &= \frac{1}{2^{j+1}}\sum_{s\in B_j}\left[
        (\Gamma_j(s)+\mu_j)^p+(2\sqrt{\mu_j H_j})^p
    \right]\\
    &= \frac{1}{2^{j+1}}\sum_{s\in B_j}
    \left[
        \sum_{r=0}^{p}\binom{p}{r}\Gamma_j(s)^{p-r}\mu_j^r + 2^{p}(\mu_j H_j)^{p/2}
    \right]\\
    &= \frac{1}{2}\left[
        \sum_{r=0}^{p}\binom{p}{r}\left(\frac{1}{2^j}\sum_{s\in B_j}\Gamma_j(s)^{p-r}\right)\mu_j^r
    \right]+ 2^{p-1}(\mu_j H_j)^{p/2}\\
    &= \frac{1}{2}\sum_{r=0}^p\binom{p}{r}R_j^{(p-r)}\mu_j^r+2^{p-1}(\mu_j H_j)^{p/2}.\label{eq350}
\end{align}

We denote 
\begin{align}
    \delta = \max\{\lambda_0,\mu_0,\mu_1,...\},
\end{align}
and we first show that $H_j=O(\delta)$, because 
\begin{align}
    H_{j+1}&=R_{j+1}^{(1)}=\frac12\left(R_j^{(1)}+R_j^{(0)}\mu_j\right)+\sqrt{\mu_j H_j}\\
    &= \frac12\left(\sqrt{H_j}+\sqrt{\mu_j}\right)^2
\end{align}
and $H_0 = \lambda_0\le \delta$
therefore if we define $h_j\equiv\sqrt{H_j/\delta}$, 
we can write 
\begin{align}
    h_{j+1} = \sqrt{\frac{H_{j+1}}{\delta}}=\frac{1}{\sqrt{2}}\left(\sqrt{\frac{H_j}{\delta}}+\sqrt{\frac{\mu_j}{\delta}}\right)
    \le\frac{1}{\sqrt{2}}\left(\sqrt{\frac{H_j}{\delta}}+1\right) = \frac{h_j+1}{\sqrt{2}},
\end{align}
which indicates
\begin{align}
    h_j\le h_*=1+\sqrt{2}, \quad \forall j,\label{eq357}
\end{align}
where $h_*$ is the fixed point of the equation $\sqrt{2}h_{j+1}=h_j+1$. 
The initial value of $h_j$ is $h_0$, which satisfies
\begin{align}
    h_0 =\sqrt{\frac{H_0}{\delta}}\le 1\le 1+\sqrt{2}.
\end{align}
So Eq.~\eqref{eq357} always holds and so 
\begin{align}
    H_j\le (3+2\sqrt{2})\delta,
\end{align} 
which is $H_j=O(\delta)$. 

Now we try to prove that $R_j^{(p)}=O(\delta^p)$ and we will use the mathematical induction method. 
Assuming for $q\le p-1$, we already have 
\begin{align}
    R_j^{(q)}\le C_q\delta ^q,\quad \forall j, 
\end{align}
and we begin from Eq.~\eqref{eq350}, 
\begin{align}
    R_{j+1}^{(p)} &= \frac{1}{2}\sum_{r=0}^p\binom{p}{r}R_j^{(p-r)}\mu_j^r+2^{p-1}(\mu_j H_j)^{p/2}\\
    &= \frac12 R_j^{(p)} + \frac12\sum_{r=1}^{p}\binom{p}{r}\mu_j^r R_j^{(p-r)} + 2^{p-1}(\mu_j H_j)^{p/2}\\
    &\le  \frac12 R_j^{(p)} + \frac12\sum_{r=1}^{p}\binom{p}{r}\mu_j^r C_{p-r}\delta^{p-r} + 2^{p-1}(C_1\mu_j\delta)^{p/2}\\
    &\le \frac12 R_j^{(p)}+\frac12\sum_{r=1}^{p}\binom{p}{r}C_{p-r}\delta^{p} + 2^{p-1} C_1^{p/2}\delta^p\\
    &= \frac12 R_j^{(p)} + \frac12\left[\sum_{r=1}^p\binom{p}{r}C_{p-r}+2^p C_1^{p/2}\right]\delta^p
\end{align}
which is 
\begin{align}
    R_{j+1}^{(p)}\le \frac{1}{2}R_j^{(p)}+D_p\delta^p
\end{align}
where $D_p=\frac12\left[\sum_{r=1}^p\binom{p}{r}C_{p-r}+2^p C_1^{p/2}\right]$. 
Therefore we have 
\begin{align}
    R_j^{(p)}&\le\frac{1}{2} R_{j-1}^{(p)} + D_p\delta^p\\
    &\le \frac12\left(\frac12 R_{j-2}^{(p)}+D_p\delta^p\right)+D_p\delta^p\\
    &...\\
    &\le \frac{1}{2^j}R_0^{(p)} + D_p\delta_p\left(\sum_{m=0}^{j-1}\frac{1}{2^m}\right)\\
    &\le \frac{1}{2^j}R_0^{(p)}+2D_p\delta^p
\end{align}
and because 
\begin{align}
    R_0^{(p)}=\lambda_0^p\le \delta ^p
\end{align}
so 
\begin{align}
    R_j^{(p)}\le (1+2D_p)\delta^p
\end{align}
which is 
\begin{align}
    R_j^{(p)}=O(\delta^p),\quad \forall j. 
\end{align}

Because we have already $R_j^{(1)}=H_j=O(\delta)$ and we have $R_j^{(0)}=1$, so according to 
the mathematical induction, we have 
\begin{align}
    R_j^{(p)}=O(\delta^p),\quad \forall p,j\in \mathbb{Z}.
\end{align}

Finally, we consider 
\begin{align}
    A_j(s,k)=\sum_{x\in B_j}(-1)^{k\cdot x}g_j(x)g_j(x\oplus s),
\end{align}
then we can write 
\begin{align}
    F_j^{(n)} = \zeta_n = \frac{1}{2^j}\sum_{s,k\in B_j}|A_j(s,k)|^{2n}.
\end{align}
We now derive the recurrence relation for $F_j^{(n)}$, we denote $s\to \sigma s\in\{0s,1s\}$ and 
$k\to\tau k\in\{0k,1k\}$. 
When $\sigma=0$, 
\begin{align}
    A_{j+1}(0s,\tau k)=A_j(s,k)+(-1)^{\tau}\mu_j\delta_{k,0}, 
\end{align}
so this part gives 
\begin{align}
    F_{j+1}^{(n)} (\sigma=0) &= F_j^{(n)} +\frac{1}{2^{j+1}}
    \sum_s\left[
        (\Gamma_j(s)+\mu_j)^{2n} + (\Gamma_j(s)-\mu_j)^{2n} - 2\Gamma_j(s)^{2n}
    \right]\\
    &= F_j^{(n)} + \sum_{\ell=1}^n\binom{2n}{2\ell} \mu_j^{2\ell}R_j^{(2n-2\ell)}, 
\end{align}
and for $\sigma=1$, we first define 
\begin{align}
    G_j(k)=\sum_{x\in B_j}(-1)^{k\cdot x}g_j(x),
\end{align}
and the calculation yields
\begin{align}
    A_{j+1}(1s,\tau k) = q_jG_j(k)[1+(-1)^{\tau+k\cdot s}], 
\end{align}
where we denote $q_j=\sqrt{\mu_j/2^j}$, and 
we further define 
\begin{align}
    Q_j^{(n)} = 2^{-jn}\sum_k|G_j(k)|^{2n}, 
\end{align}
and the contribution of this part is 
\begin{align}
    F_{j+1}^{(n)} (\sigma=1) = 2^{2n-1}\mu_j^n Q_j^{(n)}.
\end{align}
In summary we have 
\begin{align}
    F_{j+1}^{(n)} = F_j^{(n)} + \sum_{\ell=1}^n\binom{2n}{2\ell}\mu_j^{2\ell}R_j^{(2n-2\ell)}
    +2^{2n-1}\mu_j^n Q_j^{(n)}.
\end{align}

We now try to prove that $Q_j^{(n)}=O(\delta^n)$.
Denoting 
\begin{align}
    p_k=\frac{|G_j(k)|^2}{2^j}
\end{align}
we can write 
\begin{align}
    Q_j^{(n)} = \sum_k p_k^n
\end{align}
bacause $g_j(x)\ge 0$ so 
\begin{align}
    |G_j(k)|\le G_j(0)\Rightarrow p_k\le p_0
\end{align}
and 
\begin{align}
    p_0 = \frac{G_j(0)^2}{2^j}=H_j
\end{align}
so 
\begin{align}
    p_k\le H_j.
\end{align}

In the other side, due to 
\begin{align}
    \Gamma_j(s)=\sum_x g_j(x) g_j(x\oplus s)
\end{align}
we have 
\begin{align}
    \sum_s(-1)^{k\cdot s}\Gamma_j(s) &= \sum_{s,x}(-1)^{k\cdot s} g_j(x) g_j(x\oplus s)\\
    &= \sum_{x,y} (-1)^{k\cdot(x\oplus y)} g_j(x)g_j(y)\\
    &= G_j(k)^2
\end{align}
and we further have 
\begin{align}
    \sum_k G_j(k)^4 = 2^j\sum_s\Gamma_j(s)^2
\end{align}
so we have 
\begin{align}
    Q_j^{(2)} = 2^{-2j}\sum_k G_j(k)^4=2^{-j}\sum_s\Gamma_j(s)^2=R_j^{(2)}
\end{align}
which is 
\begin{align}
    Q_j^{(2)} = R_j
\end{align}
therefore for general $n\ge 2$, we have 
\begin{align}
    Q_j^{(n)}=\sum_k p_k^n = \sum_k p_k^{n-2} p_k^2\le \sum_k H_j^{n-2} p_k^2 = H_j^{n-2}R_j^{(2)} = O(\delta^n). 
\end{align}

Now we go back to $F_{j}^{(n)}$, we have 
\begin{align}
    F_{j+1}^{(n)} - F_j^{(n)} &= \sum_{\ell=1}^n\binom{2n}{2\ell}\mu_j^{2\ell}R_j^{2n-2\ell}
    +2^{2n-1}\mu_j^n Q_j^{(n)}\\
    &\le \underbrace{
        \left[\left(\sum_{\ell=1}^{n}\binom{2n}{2\ell}C_j\right) + C_{Q}\right]
    }_{C_n}
        \mu_j\delta^{2n-1}
    =C_n\mu_j\delta^{2n-1}.
\end{align}
Because we have 
\begin{align}
    F_0^{(n)} = \lambda_0^{2n}\le \lambda_0\delta^{2n-1},
\end{align}
so sum over all shells we have 
\begin{align}
    F_{J}^{(n)}\le C_n\underbrace{\left(\lambda_0+\sum_{j<J}\mu_j\right)}_{Z_J}\delta^{2n-1}
\end{align}
where we define $Z_J$ as the total weight. 

The lower bound is much simpler, when $\ell=n$, $R_j^{(0)}=1$, 
so we have 
\begin{align}
    F_{j+1}^{(n)}\ge F_j^{(n)}+\mu_j^{2n}
\end{align}
which gives
\begin{align}
    F_J^{(n)}\ge\lambda_0^{2n}+\sum_{j<J}\mu_j^{2n}.
\end{align}

Therefore for any shell-flat spectrum we get 
\begin{align}
    \lambda_0^{2n}+\sum_{j<J} \mu_j^{2n}\le F^{(n)}_J \le C_n Z_J \delta^{2n-1}, \quad \forall J,
\end{align}
which is also 
\begin{align}
    \lambda_0^{2n}+\sum_j \mu_j^{2n}\le \zeta_n \le C_n Z \delta^{2n-1}. 
\end{align}

\subsection{Sandwich construction}
From real Schmidt spectrum 
\begin{align}
    \lambda_0\ge\lambda_1\ge...,
\end{align}
we still use $B_j$ and $S_j$ notation and define 
\begin{align}
    \mu_j^+ & =2^j\lambda_{2^j}\\
    \mu_j^- & =2^j\lambda_{2^{j+1}-1}
\end{align}
so for every $x\in S_j$ we have 
\begin{align}
    \lambda_{2^{j}}\ge \lambda_x\ge\lambda_{2^{j+1}-1}.
\end{align}
We use $\overline{\lambda}$ and $\underline{\lambda}$ to pinch the real spectrum $\lambda$,
\begin{align}
    \underline{\lambda}_x\le \lambda_x\le\overline{\lambda}_x,\quad \forall x
\end{align}
where $\underline{\lambda}_x=\mu_j^-/2^j$ for $x\in S_j$ and $\overline{\lambda}_x=\mu_j^+/2^j$ for $x\in S_j$. 
From 
\begin{align}
    F^{(n)}[\lambda] = \sum_s \sum_{\substack{x_1,x_2,...,x_{2n}\\x_1\oplus...\oplus x_{2n}=0}}\prod_{r=1}^{2n}\sqrt{\lambda_{x_r}\lambda_{x_r\oplus s}}
\end{align}
we can easily have 
\begin{align}
    F^{(n)}[\underline{\lambda}]\le F^{(n)}[\lambda] \le F^{(n)}[\overline{\lambda}]. 
\end{align}

Define 
\begin{align}
    \delta_{+}=\max\{\lambda_0,\sup_j \mu_j^+\}
\end{align}
so for $\overline{\lambda}$, the total weight is 
\begin{align}
    Z_+ = \lambda_0+\sum_j\mu_j^+
\end{align}
becasue we have 
\begin{align}
    2^j\lambda_{2^j}\le 2\sum_{x=2^{j-1}}^{2^j-1}\lambda_x
\end{align}
so 
\begin{align}
    \sum_{j\ge 1}\mu_j^+\le 2(1-\lambda_0)
\end{align}
therefore ($\mu_0^+=\lambda_1\le\lambda_0$)
\begin{align}
    Z_+=\lambda_0+\mu_0^++\sum_{j\ge1}\mu_j^+ \le 2-\lambda_0+\lambda_1\le 2\Rightarrow Z_+=O(1)
\end{align}
so we can write 
\begin{align}
    F^{(n)}[\lambda] \le C_n\delta_+^{2n-1}
\end{align}
and for lower bound we have 
\begin{align}
    F^{(n)}[\lambda] \ge \lambda_0^{2n}+\sum_j(\mu_j^-)^{2n}
\end{align}
and put together 
\begin{align}
    \lambda_0^{2n}+\sum_j(\mu_j^-)^{2n} \le F^{(n)}[\lambda] \le C_n\delta_+^{2n-1}. 
\end{align}

\subsection{Scaling analysis of Calabrese--Lefevre spectrum}
The Calabrese--Lefevre spectrum tells us that~\cite{Calabrese2008}
\begin{align}
    n(\lambda) = I_0\left(
        2\sqrt{b\ln\frac{\lambda_{\max}}{\lambda}}
    \right)
\end{align}
where $b = -\ln \lambda_{\max}$. Ideally, we should have 
\begin{align}
    n(\lambda_x)=x+1.
\end{align}
We denote 
\begin{align}
    z_x\equiv2\sqrt{b\ln\frac{\lambda_{\max}}{\lambda_x}},
\end{align}
we can write $n(\lambda_x)=I_0(z_x)=x+1$
and also we obtain 
\begin{align}
    z_x=2\sqrt{b\ln\frac{\lambda_{\max}}{\lambda_x}}&\Rightarrow \frac{z_x^2}{4b} = \ln\frac{\lambda_{\max}}{\lambda_x}\\
    &\Rightarrow \lambda_x=\lambda_{\max} \exp\left(-\frac{z_x^2}{4b}\right)=\exp\left(-b-\frac{z_x^2}{4b}\right). 
\end{align}

For simplicity, we define 
\begin{align}
    U_b(x):= (x+1)\lambda_x = I_0(z_x)e^{-b-z_x^2/(4b)}, 
\end{align}
for sufficient large $z$, we have 
\begin{align}
    I_0(z)\sim\frac{e^z}{\sqrt{2\pi z}}\left(
        1+\frac{1}{8z}+\frac{9}{128z^2}+\frac{75}{1024 z^3}+...
    \right), 
\end{align}
so for large $z_x$ we have 
\begin{align}
    U_b(x)\sim \frac{1}{\sqrt{z_x}}\exp\left(
        z_x-b-\frac{z_x^2}{4b}
    \right)=\frac{1}{\sqrt{z_x}}\exp\left[
        -\frac{(z_x-2b)^2}{4b}
    \right].
\end{align}
When $0\le z_x\le b$, we have 
\begin{align}
    I_0(z)\le e^z
    \Rightarrow
    U_b(x) \le \exp\left(-b+z_x-\frac{z_x^2}{4b}\right)\le \exp\left(-\frac{b}{4}\right).
\end{align}
When $z_x\ge b$, if $b$ is large enough,
\begin{align}
    U_b(x)\le C_I\frac{1}{\sqrt{z_x}}\exp\left[
        -\frac{(z_x-2b)^2}{4b}
    \right],
\end{align}
where $C_I$ is a constant, and because $z_x\ge b$, we have $1/\sqrt{z_x}\le 1/\sqrt{b}$, and the exponential term is always smaller 
than 1, so we have 
\begin{align}
    U_b(x)\le C_I b^{-1/2},
\end{align}
therefore we finally obtain 
\begin{align}
    \sup_x U_b(x)\le C_I b^{-1/2}. 
\end{align}

Because we have $\mu_j^+=2^j\lambda_{2^j}$, so $U_b(2^j)=(2^j+1)\lambda_{2^j}$, and further we have 
\begin{align}
    \mu_j^{+}=\frac{2^j}{2^j+1} U_b(2^j)\le U_b(2^j),
\end{align}
therefore we also obtain 
\begin{align}
    \sup_j \mu_j^{+}\le C b^{-1/2},
\end{align}
and because $\lambda_0=e^{-b}\le C b^{-1/2}$, so in summary we have 
\begin{align}
    \delta_+=O(b^{-1/2}).\label{eq203}
\end{align}
By using the result above, we directly have 
\begin{align}
    F^{(n)}\le C Z_+ \delta_+^{2n-1}\Rightarrow F^{(n)}\le C b^{-n+\frac12} \Rightarrow F^{(n)}=O(b^{-n+\frac12}). \label{upperbound}
\end{align}

Now we further use 
\begin{align}
    F^{(n)}\ge \lambda_0^{2n}+\sum_j (\mu_j^-)^{2n}.
\end{align}
From $\mu_j^-=2^j\lambda_{2^{j+1}-1}$, we denote $y_j = z_{2^{j+1}-1}$
so we have $I_0(y_j)=2^{j+1}$ and further 
\begin{align}
    \lambda_{2^{j+1}-1}=\exp\left(-b-\frac{y_j^2}{4b}\right)\Rightarrow
    \mu_j^-=2^j \exp\left(-b-\frac{y_j^2}{4b}\right)
\end{align}
by using $2^j=\frac12 I_0(y_j)$, we obtain 
\begin{align}
    \mu_j^- = \frac{1}{2}I_0(y_j)\exp\left(-b-\frac{y_j^2}{4b}\right)
\end{align}
and also at large $y_j$, we get
\begin{align}
    \mu_j^-\sim\frac{1}{\sqrt{y_j}}\exp\left[-\frac{(y_j-2b)^2}{4b}\right].
\end{align}

For a fixed constant $A>0$, we define 
\begin{align}
    \mathcal{J}_A(b)=\{j:|y_j-2b|\le A\sqrt{b}\},
\end{align}
and for these $j\in \mathcal{J}_A(b)$, if $b$ is large enough, 
\begin{align}
    2b-A\sqrt{b}\le y_j\le 2b+A\sqrt{b}\Rightarrow b\le y_j \le 3b,
\end{align}
so we get 
\begin{align}
    \frac{1}{\sqrt{y_j}}\ge\frac{1}{\sqrt{3b}},
\end{align}
and we also have 
\begin{align}
    \frac{(y_j-2b)^2}{4b}\le \frac{A^2}{4} \Rightarrow \exp\left[-\frac{(y_j-2b)^2}{4b}\right]\ge e^{-A^2/4},
\end{align}
therefore we obtain 
\begin{align}
    \mu_j^-\ge C b^{-1/2},\quad j\in \mathcal{J}_A(b),
\end{align}
where $C$ is some constant. 

Now we need to estimate the size of $\mathcal{J}_A(b)$. 
From $I_0(y_j)=2^{j+1}$, we have 
\begin{align}
    (j+1)\ln 2 = \ln I_0(y_j) = y_j-\frac{1}{2}\ln 2\pi y_j + \ln(1+\frac{1}{8y_j}+...)
\end{align}
the window length is about ($y_{\pm}=2b\pm A\sqrt{b}$)
\begin{align}
    \frac{\ln I_0(y_+)-\ln I_0(y_-)}{\ln 2} &= \frac{1}{\ln 2}\left[
        (y_+-y_-)-\frac12\ln\frac{y_+}{y_-}+ \ln\left(\frac{1+\frac{1}{8y_+}+...}{1+\frac{1}{8y_-}+...}\right)
    \right]\\
    &= \frac{1}{\ln 2}\left[
        2A\sqrt{b} - \frac{1}{2}\ln\frac{2b+A\sqrt{b}}{2b-A\sqrt{b}}
        +\ln\left(\frac{1+\frac{1}{8(2b+A\sqrt{b})}+...}{1+\frac{1}{8(2b-A\sqrt{b})}+...}\right)
    \right]\\
    &= \frac{1}{\ln 2}\left(
        2A\sqrt{b} - \frac{A}{2}\frac{1}{\sqrt{b}}+ ... 
    \right),
\end{align}
so in general we have 
\begin{align}
    |\mathcal{J}_A(b)|=\frac{2A}{\ln 2}\sqrt{b} + O(1), 
\end{align}
which is also 
\begin{align}
    |\mathcal{J}_A(b)|=\Theta(\sqrt{b}). 
\end{align}

For each $j\in\mathcal{J}_A(b)$, we have 
\begin{align}
    \mu_j^-\ge C_1 b^{-1/2}
\end{align}
and there are $|\mathcal{J}_A(b)|$ number of such $j$, 
\begin{align}
    |\mathcal{J}_A(b)|\ge C_2\sqrt{b}
\end{align}
so 
\begin{align}
    F^{(n)}\ge\sum_j(\mu_j^-)^{2n}\ge\sum_{j\in \mathcal{J}_A}(\mu_j^-)^{2n} = C_2\sqrt{b}(C_1 b^{-1/2})^{2n}=C b^{-n+\frac12},
\end{align}
therefore we also have 
\begin{align}
    F^{(n)}\ge C b^{-n+\frac12} \label{lowerbound}
\end{align}
for some constant $C$. 

From Eqs.~\eqref{upperbound} and \eqref{lowerbound}, we have 
\begin{align}
    c b^{-n+\frac12}\le F^{(n)} \le C b^{-n+\frac12}\Rightarrow F^{(n)}\sim b^{-n+\frac12}.
\end{align}
for some constant $c$ and $C$. 
Therefore, the nonlocal SRE 
\begin{align}
    M_{\alpha}^{\mathrm{Sch}}(L\to+\infty) = \frac{1}{1-\alpha}\log_2 F^{(\alpha)}\sim \frac{2\alpha-1}{2\alpha-2}\log_2 b,
\end{align}
and in CFT we have 
\begin{align}
    b = \frac{c}{6}\ln L + O(1),
\end{align}
so finally we have 
\begin{align}
     M_{\alpha}^{\mathrm{Sch}}(L\to+\infty) \sim \frac{2\alpha-1}{2\alpha-2}\log_2\left[\frac{c}{6}\ln L+O(1)\right]\sim\frac{2\alpha-1}{2\alpha-2}\log_2\ln L+O(1). 
\end{align}
where $L$ is the effective size of the system, or subsystem size in this context. 

\subsection{Scaling analysis of algebraically decaying spectrum at $p=1$}
Now we study algebraically decaying spectrum given by 
\begin{align}
    \lambda_x=\frac{1}{H_N}\frac{1}{1+x}, \quad x=0,1,...,N
\end{align}
where 
\begin{align}
    H_N = \sum_{x=0}^{N-1}\frac{1}{1+x}=\sum_{k=1}^N\frac{1}{k}\sim \ln N + \gamma + o(1),
\end{align}
and $N=2^m$. 

We still use $\mu_j^+=2^j\lambda_{2^j}$ and $\mu_j^-=2^j\lambda_{2^{j+1}-1}$. Luckily, in this spectrum, these two 
quantities can be exactly evaluated, 
\begin{align}
    \mu_j^+=\frac{2^j}{2^j+1}\frac{1}{H_N}=\Theta(H_N^{-1}), \label{eq256}
\end{align}
and also 
\begin{align}
    \mu_j^-=2^j\lambda_{2^{j+1}-1}=\frac{2^j}{2^{j+1}}\frac{1}{H_N}=\frac{1}{2H_N}=\Theta(H_N^{-1}). 
\end{align}

Therefore, use the abovementioned conclusion,  for lower bound we have 
\begin{align}
    \zeta_n\ge \lambda_0^{2n}+\sum_{j=0}^{m-1}(\mu_j^-)^{2n}=\frac{1}{H_N^{2n}}+\frac{m}{2^{2n}H_N^{2n}},
\end{align}
and becasue $m=\log_2 N\sim H_N/\ln 2 + o(1)$, we can write 
\begin{align}
    \zeta_n\ge H_{N}^{-2n}+\frac{2^{-2n}}{\ln 2}H_N^{-2n+1}\ge C_n H_N^{-(2n-1)}
\end{align}
for some constant $C_n$. 

And from Eq.~\eqref{eq256}, we directly have 
\begin{align}
    \delta^+=\frac{1}{H_N}, 
\end{align}
and 
\begin{align}
    Z_+=\frac{1}{H_N}+\frac{1}{H_N}\sum_{j=0}^{m-1}\frac{2^j}{2^j +1}\le \frac{m+1}{H_N}\sim\frac{m+1}{m \ln 2} + O(1)
\end{align}
so when $m\to +\infty$, $Z_+=O(1)$, consequently 
\begin{align}
    \zeta_n\le C_n H_N^{-(2n-1)}
\end{align}
for some constant $C_n$. 

In summary we then obtain 
\begin{align}
    c_n H_N^{-(2n-1)}\le \zeta_n\le C_n H_N^{-(2n-1)}, 
\end{align}
which means 
\begin{align}
    \zeta_n\sim \Theta\left(H_N^{-(2n-1)}\right),
\end{align}
so the scaling of nonlocal SRE is 
\begin{align}
    M_{\alpha}^{\mathrm{Sch}}=\frac{1}{1-\alpha}\log_2\zeta_{\alpha}\sim \frac{2\alpha-1}{\alpha-1}\log_2 H_N\sim 
    \frac{2\alpha-1}{\alpha-1}\log_2\ln N. 
\end{align}

\section{Numerical details}
\subsection{Entanglement spectrum from corner transfer matrix}
We consider the TFIM in disorder phase, and study the half-chain reduced density matrix, which is given by 
\begin{align}
    \rho_A = \frac{1}{Z}\exp\left(
        -\sum_{j=0}^{+\infty}\epsilon_j\hat{c}_j^{\dagger}\hat{c}_j
    \right),
\end{align}
and when $h>1$, the entanglement energies are given by 
\begin{align}
    \epsilon_j = (2j+1)\epsilon,\quad j = 0,1,2,...
\end{align}
when $\epsilon\to0$, we are approaching critical point. 
The entanglement eigenvalues are 
\begin{align}
    \lambda_{\{n_j\}} = \frac{1}{Z}\exp\left[
        -\epsilon\sum_{j=0}^{+\infty}(2j+1)n_j
    \right], \quad n_j\in \{0,1\}
\end{align}
and the normalization factor is 
\begin{align}
    Z(\epsilon)=\prod_{j=0}^{+\infty}\left(1 + e^{-(2j+1)\epsilon}\right).
\end{align}
The largest eigenvalue corresponds to $n_j=0,\forall j$, so we have 
\begin{align}
    \lambda_{0}=\frac{1}{Z},
\end{align}
and we can define 
\begin{align}
    b = -\ln\lambda_0 = \ln Z,
\end{align}
and we naturally have 
\begin{align}
    b(\epsilon)=\sum_{j=0}^{+\infty}\ln\left[1+e^{-(2j+1)\epsilon}\right].
\end{align}

Because we focus on the case that $b$ is large, and this corresponds to the case $\epsilon<<1$, and when $\epsilon <<1$
\begin{align}
    b(\epsilon)=\sum_{j=0}^{+\infty}\ln\left[1+e^{-(2j+1)\epsilon}\right]
    \simeq \frac{1}{2\epsilon}\int_0^{+\infty}\mathrm{d}x\ \ln(1+e^{-x}) = \frac{\pi^2}{24\epsilon},
\end{align}
so in this limit we have 
\begin{align}
    \epsilon\simeq\frac{\pi^2}{24b}. 
\end{align}

Now we try to figure out the ordering of the spectrum. 
Define the integer $q$, 
\begin{align}
    q = \sum_{j=0}^{+\infty}(2j+1)n_j,
\end{align}
then all eigenvalues depend on $q$, 
\begin{align}
    \lambda(q)=e^{-b-\epsilon q},
\end{align}
so 
\begin{align}
    q_1<q_2\Rightarrow \lambda(q_1)>\lambda(q_2).
\end{align}

So we only need to deal with the ordering of $q$. But there is a degeneracy problem. 
For a fixed $q$, we denote there are $d_q$ corresponding occupation configurations. 
For every odd integer $1, 3, 5, 7, ...$, it can only be choosed 0 or 1 time, because $n_j=0,1$. 
Then the generating function is 
\begin{align}
    \sum_{q=0}^{+\infty} d_q t^q = \sum_{j=0}^{+\infty}(1+t^{2j+1}).
\end{align}
We further define cumulative degeneracy 
\begin{align}
    D_q=\sum_{s=0}^q d_s, 
\end{align}
then we can find only one $q=q(x)$ satisfying
\begin{align}
    D_{q-1}\le x< D_q
\end{align}
so 
\begin{align}
    \lambda_x=e^{-b-\epsilon q(x)}. 
\end{align}
This corner transfer matrix (CTM) solution of the entanglement spectrum of Ising model is given by Ref.~\cite{Peschel1999}, to which we refer for further details. 

\subsection{Monte Carlo sampling}
Denote $f(x)=\sqrt{\lambda_x}$, the $\zeta_2$ is defined as 
\begin{align}
    \zeta_2^{\mathrm{Sch}} = 2^{-M_2^{\mathrm{Sch}}}=\sum_{x,a,c,d\in \mathbb{F}_2^m}\prod_{\omega\in \mathbb{F}_2^3}
    f(x\oplus \omega_1 a\oplus\omega_2 c\oplus \omega_3 d),
\end{align}
if we define the cube configuration
\begin{align}
    C=(x,a,c,d),
\end{align}
with its eight vertices, 
\begin{align}
    v_1&=x,\\
    v_2&=x\oplus a,\\
    v_3&=x\oplus c,\\
    v_4&=x\oplus d,\\
    v_5&=x\oplus a\oplus c,\\
    v_6&=x\oplus a\oplus d,\\
    v_7&=x\oplus c\oplus d,\\
    v_8&=x\oplus a\oplus c\oplus d,
\end{align}
and we define 
\begin{align}
    W(C)=\prod_{i=1}^{8}\sqrt{\lambda_{v_i}}=\prod_{i=1}^{8}f(v_i),
\end{align}
so we have 
\begin{align}
    \zeta_2^{\mathrm{Sch}}=\sum_C W(C). 
\end{align}

Define the auxiliary energy 
\begin{align}
    E(C)=-\ln W(C), 
\end{align}
then 
\begin{align}
    E(C) = -\frac12\sum_{i=1}^{8}\ln\lambda_{v_i},
\end{align}
and for CTM spectrum we have 
\begin{align}
    -\ln \lambda_x = b_{\mathrm{eff}} + \epsilon q(x), 
\end{align}
so 
\begin{align}
    E(C) = \frac{1}{2}\sum_{i=1}^8b_{\mathrm{eff}} + \epsilon q(x) = 4b_{\mathrm{eff}}+\frac{\epsilon}{2}\sum_{\omega\in \mathbb{F}_2^3}q(v_i). 
\end{align}

Now we introduce the thermodynamic integration. Define an inverse temperature $\beta\in[0,1]$, 
and a new partition function
\begin{align}
    Z(\beta)=\sum_{C}e^{-\beta E(C)}=\sum_C W(C)^{\beta},
\end{align}
and we have 
\begin{align}
    Z(\beta=1)&=\zeta_2^{\mathrm{Sch}},\\
    Z(\beta=0)&=2^{4m},
\end{align}
where $N=2^m$ and $N$ is the size of the spectrum. 

\begin{figure*}
    \includegraphics[width=0.75\linewidth]{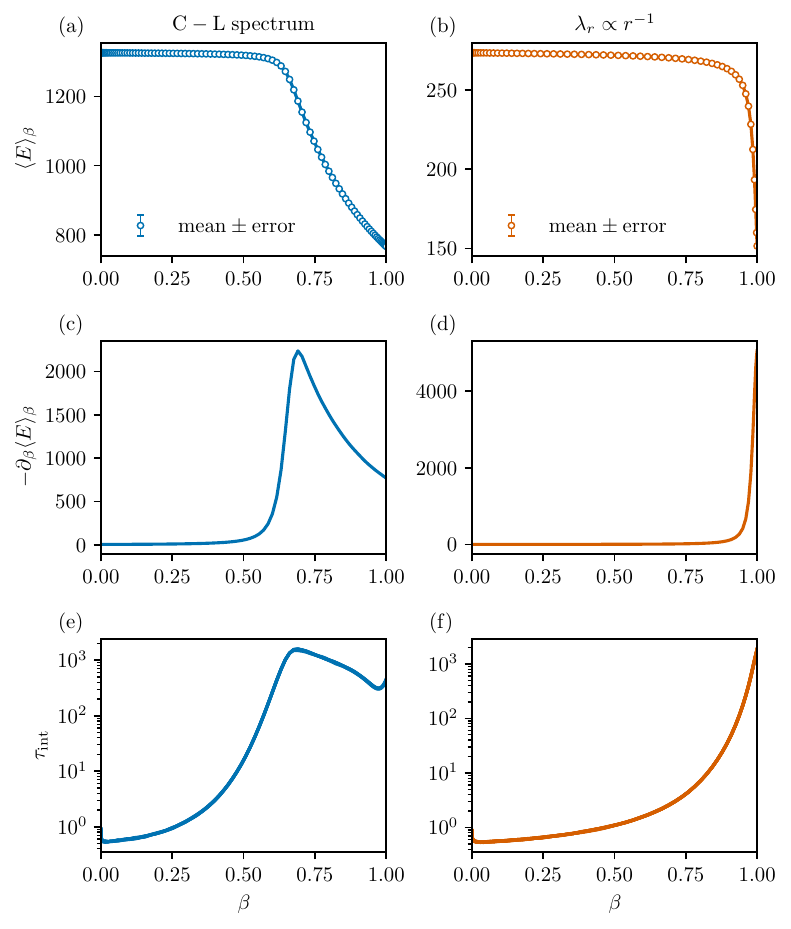}
    \caption{
    (a,c,d) Energy $\braket{E}_{\beta}$, its derivative $-\partial_{\beta}\braket{E}_{\beta}$ and integrated autocorrelation time $\tau_{\mathrm{int}}$ when sampling C--L spectrum at $b=96$. 
    (b,d,f) The same quantities for algebratically decay spectrum $\lambda_r\propto r^{-1}$ at $L = 94$.  
    \label{fig2}
    }
\end{figure*}

We have 
\begin{align}
    \frac{\mathrm{d}}{\mathrm{d}\beta}\ln Z(\beta) = \frac{1}{Z(\beta)}\sum_C[-E(C)]e^{-\beta E(C)}=-\braket{E}_{\beta}, 
\end{align}
where $P_{\beta}(C)=e^{-\beta E(C)}/Z(\beta)$ is the auxiliary Boltzmann distribution. 
And now we have 
\begin{align}
    \ln Z(1)-\ln Z(0)=-\int_0^1\mathrm{d}\beta\braket{E}_{\beta},
\end{align}
which is also 
\begin{align}
    \ln \zeta_2^{\mathrm{Sch}}=4m\ln 2-\int_0^1\mathrm{d}\beta \braket{E}_{\beta}, 
\end{align}
so finally we have 
\begin{align}
    M_2^{\mathrm{Sch}} = -\frac{\ln \zeta_2^{\mathrm{Sch}}}{\ln 2}
    =\frac{1}{\ln 2}\left[\int_0^1\mathrm{d}\beta\braket{E}_{\beta}-4m\ln 2\right].
\end{align}
Therefore we only need to use MC to calculate $\braket{E}_{\beta}$ and evaluate the integral. 
This Monte Carlo sampling method was introduced in Ref.~\cite{Sierant2026}, to which we refer for further details. 
It is interesting to note that there is a potential phase transition for C--L spectrum at around $\beta=0.63$, as shown in Fig.~\ref{fig2}. And the $p=1$ power-law decaying spectrum behaves differently, the derivative peaks at $\beta=1$. 

\end{document}